\documentclass[a4paper,11pt]{article}

\usepackage{amsmath,amsthm,amsbsy,amssymb,amsfonts,graphicx,mathrsfs,bm,cite,color}
\usepackage{hyperref}
\usepackage[utf8]{inputenc}

\makeatletter
\catcode`\@=11
\@addtoreset{equation}{section}

\makeatother

\makeatletter

\@addtoreset{table}{section}
\makeatother

\renewcommand{\thefootnote}{\arabic{footnote}}

\newcommand{\Exp}[1]{\operatorname{e}^{#1}}

\newcommand{\abs}[1]{\lvert {#1} \rvert}
\newcommand{\rmd}{{\mathrm{d}}}
\newcommand{\nn}{\nonumber}
\newcommand{\Lie}{\pounds}
\newcommand{\gLie}{\hat{\pounds}}

\newcommand{\cA}{\mathcal A}

\newcommand{\cF}{\mathcal F}
\newcommand{\cH}{\mathcal H}
\newcommand{\cI}{\mathcal I}\newcommand{\cJ}{\mathcal J}
\newcommand{\cK}{\mathcal K}\newcommand{\cL}{\mathcal L}
\newcommand{\cM}{\mathcal M}

\newcommand{\cT}{\mathcal T}

\newcommand{\sfg}{\mathsf{g}}
\newcommand{\sfh}{\mathsf{h}}

\newcommand{\SL}{\text{SL}}

\newcommand{\GL}{\text{GL}}
\newcommand{\OO}{\text{O}}

\newcommand{\rmT}{\text{T}}

\newcommand{\nafla}{\mathring{\nabla}}

\theoremstyle{definition}
\newtheorem{exmp}{Example}

\allowdisplaybreaks[3]

\begin{document}

\begin{titlepage}
\renewcommand{\thefootnote}{\fnsymbol{footnote}}

\vspace*{1.0cm}

\begin{center}
\centerline{\Large\textbf{Born sigma model for branes in exceptional geometry}}%
\end{center}
\vspace{1.0cm}

\centerline{\large
{Yuho Sakatani}%
\footnote{E-mail address: \texttt{yuho@koto.kpu-m.ac.jp}}
\ \ and \ \ 
{Shozo Uehara}%
\footnote{E-mail address: \texttt{uehara@koto.kpu-m.ac.jp}}
}

\vspace{0.2cm}

\begin{center}
{\it Department of Physics, Kyoto Prefectural University of Medicine,}\\
{\it Kyoto 606-0823, Japan}
\end{center}

\vspace*{2mm}

\begin{abstract}
In double field theory, the physical space has been understood as a subspace of the doubled space. Recently, the doubled space is defined as the para-Hermitian manifold and the physical space is realized as a leaf of a foliation of the doubled space. This construction naturally introduces the fundamental 2-form, which plays an important role in a reformulation of string theory known as the Born sigma model. In this paper, we present the Born sigma model for $p$-branes in M-theory and type IIB theory by extending the fundamental 2-form into $U$-duality-covariant $(p+1)$-forms. 
\end{abstract}

\thispagestyle{empty}
\end{titlepage}

\setcounter{footnote}{0}

\newpage

\tableofcontents

\newpage

\section{Introduction}

Double Field Theory (DFT) \cite{hep-th:9302036,hep-th:9305073,hep-th:9308133,0904.4664,1006.4823} has been developed for the $T$-duality-covariant formulation of supergravity. 
This is defined on a $2d$-dimensional space called the doubled space, but in order to consistently formulate the theory, a constraint known as the section condition is required. 
Accordingly, we require that any fields depend only on at most $d$ coordinates. 
Namely, we suppose that any fields are defined on a $d$-dimensional physical space. 
The choice of the physical directions is arbitrary and it can be specified by the polarization tensor \cite{hep-th:0406102}. 
This arbitrariness in the choice of the polarization tensor can be understood as $T$-duality symmetry. 
Although the polarization tensor is an important object in the doubled space, it appears only implicitly in the conventional formulation of DFT. 
This is because we usually consider a specific class of physical spaces. 
As we review in section \ref{sec:review}, if we consider a general polarization, the polarization tensor appears explicitly, for example, in the definition of the generalized Lie derivative. 
This kind of general polarizations will be important to investigate non-trivial applications of DFT, and various aspects have been studied recently in \cite{1203.0836,1209.0152,1706.07089,1802.08180,1806.05992,1810.03953,1901.04777}. 

In this recent approach, the physical space is specified by the polarization tensor, or equivalently, an almost para-complex structure $K^I{}_J$ satisfying $(K^2)^I{}_J=\delta^I_J$ (see \cite{yano1959} for details of the para-complex structure). 
A set of $d$ eigenvectors with the eigenvalue $+1$ spans the tangent space of the physical space while a set of $d$ eigenvectors with the eigenvalue $-1$ spans the unphysical gauge orbits in the sense of \cite{1304.5946}. 
The polarization tensors $\Pi_\pm$, which pick out the physical/unphysical directions, are defined as $(\Pi_\pm)^I{}_J\equiv\frac{1}{2}\,\bigl(\delta^I_J \pm K^I{}_J\bigr)$ and the section condition is expressed as $(\Pi_-)^J{}_I\,\partial_J f(x)=0$\,. 
When $K^I{}_J$ satisfies the integrability condition, it is called a para-complex structure, and it allows us to find a local coordinate system $(x^I)=(x^m,\,\tilde{x}_m)$ such that the physical and the unphysical space are realized as $\tilde{x}_m=\text{const.}$ and $x^m=\text{const.}$, respectively.
Then, $K^I{}_J$ takes the form
\begin{align}
 (K^I{}_J) = \begin{pmatrix} \delta^m_n & 0 \\ 0 & -\delta_m^n \end{pmatrix},
\label{eq:adapted-K}
\end{align}
and the section condition becomes $\tilde{\partial}^m f(x)=0$\,. 
In this viewpoint, the section condition can be understood as a condition that any fields on the doubled space (a para-complex manifold) are para-holomorphic functions. 
Consequently, similar to the case of holomorphic functions defined on a complex plane, any fields satisfying the section condition can be consistently restricted to a half-dimensional physical subspace. 

In this recent construction of the doubled space, by further introducing the standard $\OO(d,d)$ metric $\eta_{IJ}$\,, the para-complex manifold becomes the para-Hermitian manifold. 
There, we can define a natural 2-form field $\omega_{IJ}\equiv \eta_{IK}\,K^K{}_J$\,, called the fundamental 2-form. 
In the particular coordinate system \eqref{eq:adapted-K}, this 2-form has the form
\begin{align}
 (\omega_{IJ}) = \begin{pmatrix} 0 & -\delta_m^n \\ \delta^m_n & 0 \end{pmatrix},
\end{align}
and can be interpreted as a symplectic form. 
Accordingly, the interpretation of the doubled space as a kind of phase space has been developed in \cite{1307.7080,1502.08005,1606.01829,1706.03305,1707.00312,1812.10821}, and there, the $T$-duality is interpreted as the Fourier transformation, $x^m\to p_m$ and $p_m\to -x^m$. 
The symmetry under the Fourier transformation is known as the Born reciprocity, and the doubled space equipped with a certain dynamical metric $\cH_{IJ}$ is called the Born manifold. 
Apart from DFT, the fundamental 2-form $\omega$ appears also in the string action, known as the Tseytlin action \cite{Tseytlin:1990nb,Tseytlin:1990va}
\begin{align}
 S = \frac{1}{2}\int\rmd^2\sigma\,\bigl[(\eta+\omega)_{IJ}\,\partial_\sigma X^I\,\partial_\tau X^J - \cH_{IJ}\,\partial_\sigma X^I\,\partial_\sigma X^J\bigr]\,. 
\label{eq:Tseytlin}
\end{align}
The term including $\omega$ is topological and has not been included originally in \cite{Tseytlin:1990nb,Tseytlin:1990va}, but its importance has been discussed in \cite{hep-th:9112070,hep-th:0605149,hep-th:0701080,1111.1828,1502.08005}. 
The topological term is introduced also in Hull's approach \cite{hep-th:0406102,hep-th:0605149} where the worldsheet covariance is manifest. 
More recently, a duality-covariant string action in an arbitrary curved background is provided in \cite{1307.8377}. 
Subsequently, by further adding a certain total-derivative term, a duality-covariant action
\begin{align}
 S = \frac{1}{2}\int_{\Sigma} \Bigl[\frac{1}{2}\,\cH_{IJ}(X)\,DX^I\wedge *_\gamma DX^J - \frac{1}{2}\,\omega_{IJ}\,DX^I\wedge DX^J \Bigr]\,,
\end{align}
called the gauged Born sigma model has been proposed in \cite{1910.09997}, which explicitly utilizes $\omega$. 

In this paper, we study an extension of the Born sigma model
\begin{align}
 S = \frac{1}{p+1}\int_\Sigma \Bigl[\Exp{\lambda}\frac{1}{2}\,\cH_{IJ}(X)\,DX^I\wedge *_{\gamma} DX^J -\frac{1}{2}\,\omega^{(F)}_{IJ;\cK}\, DX^I\wedge DX^J\wedge q^\cK_{(\text{brane})}\Bigr] \,,
\end{align}
for various $p$-branes in M-theory and type IIB theory. 
It terns out that this action is the same as the one studied in \cite{1607.04265,1712.10316} but the reformulation using $\omega$ makes the structure simpler and clearer. 
In addition, the rank of the $E_n$ $U$-duality group has been assumed to satisfy $n\leq 7$ in \cite{1607.04265,1712.10316}, but here it is not assumed and we can consider the full theory $n=11$\,. 
Moreover, regarding type IIB branes, only the $(p,q)$-string has been explicitly considered in \cite{1712.10316} but here we also provide the D3-brane action and the action for the $(p,q)$-five brane. 

\medskip 

This paper is organized as follows. 
In section \ref{sec:review}, we review the description of the doubled space as the para-Hermitian manifold. 
After reviewing the geometric framework, in section \ref{sec:DSM}, we explain our approach to string sigma model. 
In section \ref{sec:exceptional}, we apply a similar discussion to the case of the exceptional space, and study sigma model actions for various $p$-branes. 
In section \ref{sec:Hamiltonian}, in order to clarify the relation to Tseytlin's approach, we study the brane actions in Hamiltonian form. 
Section \ref{sec:conclusions} is devoted to the summary and discussion. 

\section{Para-Hermitial geometry for double field theory}
\label{sec:review}

In this section, we briefly review the para-Hermitian geometry from a physics point of view. 
We then explain our approach to brane actions by using the string sigma model as a prototype. 

\subsection{Para-Hermitial geometry}

In DFT, we consider a doubled space which is a smooth $2d$-dimensional manifold $M$ endowed with a metric $\eta_{IJ}$ of signature $(d,d)$. 
We may introduce the standard Christoffel symbol $\Gamma_I{}^J{}_K$ associated with $\eta_{IJ}$ and denote the covariant derivative as
\begin{align}
 \nafla_I V^J \equiv \partial_I V^J + \Gamma_I{}^J{}_K\,V^K\,. 
\end{align}
We assume that the metric $\eta_{IJ}$ is flat and always work with a local coordinate system $(x^I)=(x^m,\,\tilde{x}_m)$ where $\eta_{IJ}$ and its inverse $\eta^{IJ}$ have the form\footnote{When $\eta_{IJ}$ is not flat, one may introduce a vielbein as $\eta_{IJ} = E_I{}^A\,E_J{}^B\,\eta_{AB}$ where $\eta_{AB}$ has the same form as Eq.~\eqref{eq:etaIJ}. See, for example, \cite{1611.07978} for more details.}
\begin{align}
 (\eta_{IJ}) = \begin{pmatrix} 0 & \delta_m^n \\ \delta^m_n & 0 \end{pmatrix},\qquad 
 (\eta^{IJ}) = \begin{pmatrix} 0 & \delta^m_n \\ \delta_m^n & 0 \end{pmatrix}.
\label{eq:etaIJ}
\end{align}
In order to consistently formulate DFT, we require that any two of the fields and gauge parameters, say $f$ and $g$, satisfy the section condition
\begin{align}
 \eta^{IJ}\,\nafla_I f\,\nafla_J g =0\,.
\label{eq:SC}
\end{align}
Here, the covariant derivative $\nafla_I$ can be replaced by $\partial_I$ because we are assuming Eq.~\eqref{eq:etaIJ} where $\Gamma_I{}^J{}_K=0$.
In the following, we review that the section condition suggests us to regard the doubled space as a para-Hermitian manifold \cite{1203.0836,1209.0152,1706.07089,1802.08180,1806.05992,1810.03953,1901.04777}. 

The section condition \eqref{eq:SC} indicates that the derivatives of any fields lie in a common null subspace. 
Accordingly, we introduce a projection operator $\Pi_+$ of rank $d$ satisfying \cite{hep-th:0406102}
\begin{align}
 \Pi_+^2 = \Pi_+\,,\qquad 
 \eta^{KL}\,(\Pi_+)^I{}_K\,(\Pi_+)^J{}_L = 0\,,
\label{eq:Pip}
\end{align}
which is known as the polarization tensor. 
Then we assume that any fields $f$ satisfy
\begin{align}
 \nafla_I f = (\Pi_+)^J{}_I\,\nafla_J f\,. 
\label{eq:linear-SC}
\end{align}
We can easily check that the section condition \eqref{eq:SC} is indeed satisfied under this situation:
\begin{align}
 \eta^{IJ}\,\nafla_I f\,\nafla_J g = \eta^{IJ}\,(\Pi_+)^K{}_I\,(\Pi_+)^L{}_J\,\nafla_K f\,\nafla_L g =0\,. 
\end{align}
If we also introduce a projection operator $\Pi_- \equiv \bm{1} - \Pi_+$ onto the orthogonal directions, Eq.~\eqref{eq:linear-SC} can be also expressed as 
\begin{align}
 (\Pi_-)^J{}_I\,\nafla_J f =0\,.
\end{align}
Then using the two projectors $\Pi_\pm$\,, we can construct an almost para-complex structure (or an almost product structure)
\begin{align}
 K \equiv \Pi_+ - \Pi_-\,,\qquad K^2 =\bm{1}\,.
\end{align}

By using the polarization tensors, we define two distributions $L^*_\pm$, spanned at each point by $(\Pi_\pm)^I{}_J \,\rmd x^J$\,. 
The condition for each of the distributions to be integrable is given by
\begin{align}
 (\Pi_\mp)^{J_1}{}_{[K_1}\,(\Pi_\mp)^{J_2}{}_{K_2]}\,\partial_{J_1} (\Pi_\pm)^I{}_{J_2} = 0\,.
\end{align}
If we define 
\begin{align}
 N_\pm(V,W) \equiv \Pi_\mp [\Pi_\pm(V),\,\Pi_\pm(W)] \qquad \bigl[\Pi_\pm(V)\equiv \Pi_\pm V \equiv (\Pi_\pm)^I{}_J\,V^J\bigr]\,,
\end{align}
for arbitrary vector fields $V^I$ and $W^I$\,, the integrability conditions are equivalent to $N_\pm=0$\,. 
If both of these distributions are integrable, we can find a local coordinate system $(x^m,\,\tilde{x}_m)$ where $\Pi_\pm$ and $K$ have the form
\begin{align}
 \Pi_+ = \Pi_+^{(0)} \equiv \begin{pmatrix} \delta^m_n & 0 \\ 0 & 0 \end{pmatrix} , \quad
 \Pi_- = \Pi_-^{(0)} \equiv \begin{pmatrix} 0 & 0 \\ 0 & \delta_m^n \end{pmatrix} , \quad
 K = K^{(0)} \equiv \begin{pmatrix} \delta^m_n & 0 \\ 0 & -\delta_m^n \end{pmatrix} .
\label{eq:K0-def}
\end{align}
Such coordinates $x^m$/$\tilde{x}_m$ are called para-holomorphic/para-anti-holomorphic coordinates, where $L^*_+$/$L^*_-$ are respectively spanned by $\rmd x^m$/$\rmd\tilde{x}_m$. 
On an overlap of two such coordinate patches $(U_\alpha,\,x_{(\alpha)}^I)$ and $(U_\beta,\,x_{(\beta)}^I)$ with $U_\alpha\cap U_\beta\neq \emptyset$\,, by requiring the para-Cauchy--Riemann equation
\begin{align}
 \frac{\partial x_{(\beta)}^m}{\partial \tilde{x}_{(\alpha)n}} = 0\,,\qquad
 \frac{\partial \tilde{x}_{(\beta)m}}{\partial x_{(\alpha)}^n} = 0\,, 
\end{align}
we can consistently define the para-complex structure in both patches as \eqref{eq:K0-def} and $K$ can be globally defined over the doubled space. 
The linear section condition \eqref{eq:linear-SC}, i.e., $\tilde{\partial}^m f=0$\,, then can be interpreted as the para-holomorphicity of any fields on the doubled space. 

The Nijenhuis tensor associated with the almost para-complex structure $K$,
\begin{align}
 N_K(V,W) \equiv -K^{2}([V,\,W]) + K([K(V),\,W]+[V,\, K(W)])-[K(V),\, K(W)] \,,
\end{align}
can be expressed as
\begin{align}
 N_K(V,W) = 4\, N_+(V,W) + 4\, N_-(V,W)\,,
\end{align}
and the integrabilities of the both two distributions $L^*_\pm$ can be summarized as the integrability of $K$: $N_K=0$\,. 
By using the polarization tensors, we can also split the tangent bundle as $TM=L_+\oplus L_-$, where distributions $L_\pm$ are defined as
\begin{align}
 L_\pm = \{V \mid \Pi_\pm(V)=V\,,\ V\in TM\} \,.
\end{align}
In this paper, we always assume that $K$ is integrable ($N_K=0$),\footnote{In order to consistently formulate DFT, only the $L_+$-integrability ($N_+ =0$) or $L_-$-integrability ($N_- =0$) may be enough. At least for the string sigma model, it is known that the $L_+$- or $L_-$-integrability is enough to formulate the gauged Born sigma model \cite{1910.09997}.} and then by using the para-(anti-)holomorphic coordinates, distributions $L_+$ and $L_-$ are respectively spanned by $\partial_m$ and $\tilde{\partial}^m$\,. 
Then, $L_+$ can be identified as the tangent bundles $T\cF_+$ of a $d$-dimensional space $\cF_+$ with coordinates $x^m$\,, which we call the physical space. 
It can be realized as $\tilde{x}_m=c_m$ ($c_m$ constant) in the para-(anti-)holomorphic coordinates. 
Similarly, the unphysical gauge orbits $\cF_-$ (for which $L_-=T\cF_-$) is described as $x^m=c^m$ ($c_m$ constant). 

So far, we have constructed the doubled space as a para-complex manifold $(M,\,K)$. 
Now let us also consider the metric $\eta_{IJ}$\,. 
The standard assumption in DFT is that the metric $\eta_{IJ}$ has the form \eqref{eq:etaIJ} in the para-(anti-)holomorphic coordinates. 
Then we can easily see that Eq.~\eqref{eq:Pip} and
\begin{align}
 \eta^{KL}\,(\Pi_-)^I{}_K\,(\Pi_-)^J{}_L = 0\,,
\label{eq:Pim}
\end{align}
are satisfied. 
Eqs.~\eqref{eq:Pip} and \eqref{eq:Pim} are equivalent to
\begin{align}
 \eta_{KL}\,K^K{}_I\,K^L{}_J = -\eta_{IJ} \,.
\end{align}
When this relation is satisfied, the pair $(K,\eta)$ is called a para-Hermitian structure and the doubled space $(M, K, \eta)$ equipped with a para-Hermitian structure is called a para-Hermitian manifold. 
On an arbitrary para-Hermitian manifold, we can define a natural 2-form field
\begin{align}
 \omega_{IJ} \equiv \eta_{IK}\,K^K{}_J \,, 
\label{eq:omega-DFT}
\end{align}
which is called the fundamental 2-form. 
By definition, this satisfies
\begin{align}
 \omega_{KL}\, (\Pi_\pm)^K{}_I \, (\Pi_\pm)^L{}_J = 0\,.
\end{align}
If this 2-form is closed $\rmd \omega=0$\,, the para-Hermitian manifold is called a para-K\"ahler manifold. 
To check the closedness in our setup, let us expand $\rmd\omega$ generally as
\begin{align}
 \rmd\omega = (\rmd \omega)^{(3,0)} + (\rmd \omega)^{(2,1)} + (\rmd \omega)^{(1,2)} + (\rmd \omega)^{(0,3)} \,,
\end{align}
where
\begin{align}
\begin{split}
 (\rmd \omega)^{(3,0)}(X,\,Y,\,Z) &\equiv \rmd \omega\bigl(\Pi_+(X),\,\Pi_+(Y),\,\Pi_+(Z)\bigr) \,,
\\
 (\rmd \omega)^{(2,1)}(X,\,Y,\,Z) &\equiv \sum_{X,Y,Z}^{\text{cyclic}}\rmd \omega\bigl(\Pi_+(X),\,\Pi_+(Y),\,\Pi_-(Z)\bigr) \,,
\\
 (\rmd \omega)^{(1,2)}(X,\,Y,\,Z) &\equiv \sum_{X,Y,Z}^{\text{cyclic}}\rmd \omega\bigl(\Pi_+(X),\,\Pi_-(Y),\,\Pi_-(Z)\bigr) \,,
\\
 (\rmd \omega)^{(0,3)}(X,\,Y,\,Z) &\equiv \rmd \omega\bigl(\Pi_-(X),\,\Pi_-(Y),\,\Pi_-(Z)\bigr) \,.
\end{split}
\end{align}
By defining $(N_\pm)_{IJK} \equiv \eta\bigl(N_\pm(\partial_I ,\, \partial_J),\,\partial_K\bigr)$\,, we can easily show
\begin{align}
 (\rmd \omega)^{(3,0)}_{IJK} = 3\,(N_+)_{[IJK]} = 0\,,\qquad 
 (\rmd \omega)^{(0,3)}_{IJK} = -3\,(N_-)_{[IJK]} = 0\,.
\end{align}
Thus in the para-Hermitian case, $(\rmd \omega)^{(3,0)}$ and $(\rmd \omega)^{(0,3)}$ vanish automatically and we obtain
\begin{align}
 \rmd\omega = (\rmd \omega)^{(2,1)} + (\rmd \omega)^{(1,2)} \,.
\end{align}
If $(\rmd \omega)^{(2,1)}=0$ and $(\rmd \omega)^{(1,2)}=0$ are further satisfied, the doubled space becomes a para-K\"ahler manifold and the fundamental 2-form $\omega$ becomes a symplectic form. 

\begin{exmp}
In the conventional DFT, we assume that the doubled space is a para-K\"ahler manifold with $K=K^{(0)}$ and $\omega=\omega^{(0)}\equiv \eta\,K^{(0)}$\,. 
We can also consider a deformation of $K^{(0)}$ by performing a $B$-transformation,
\begin{align}
 K^{(0)}\to K^{(b)} = \Exp{\bm{b}}K\Exp{-\bm{b}} = \begin{pmatrix} \delta^m_n & 0 \\ 2\,b_{mn} & -\delta_m^n \end{pmatrix} ,\quad 
 \Exp{\bm{b}} \equiv \begin{pmatrix} \delta^m_n & 0 \\ b_{mn} & \delta_m^n \end{pmatrix},\quad
 \rmd b =0 \,.
\label{eq:K-b}
\end{align}
After the deformation, $L_+$ is spanned by $e_m\equiv \partial_m - b_{mn}\,\tilde{\partial}^n$ and the linear section condition becomes $\tilde{\partial}^m=0$ (see \cite{1510.06735} where this $b_{mn}$ was introduced to discuss finite transformations in DFT). 
In addition, the fundamental 2-form becomes
\begin{align}
 \omega^{(b)} \equiv \eta\,K^{(b)} = \begin{pmatrix} 2\,b_{mn} & -\delta_m^n \\ \delta^m_n & 0 \end{pmatrix} .
\end{align}
This is still a para-K\"ahler structure because $\rmd \omega^{(b)}=0$, which follows from $\rmd b=0$\,.
\end{exmp}

\begin{exmp}
Another non-trivial example is given by
\begin{align}
 \eta = \begin{pmatrix} 0 & \delta_m^n \\ \delta^m_n & 0 \end{pmatrix} ,\qquad
 K^{(\pi)} = \begin{pmatrix} \delta^m_n & 2\,\pi^{mn} \\ 0 & -\delta_m^n \end{pmatrix} ,\qquad
 \omega^{(\pi)} = \begin{pmatrix} 0 & -\delta_m^n \\ \delta^m_n & 2\,\pi^{mn} \end{pmatrix} ,
\end{align}
where $\pi^{mn}$ is a Poisson-tensor satisfying $\pi^{m[n}\,\partial_m \pi^{pq]} = 0$.
In this case, we obtain
\begin{align}
 \rmd\omega = (\rmd \omega)^{(1,2)} = \partial_p \pi^{mn}\,\bigl(\rmd x^p +\pi^{pq}\,\rmd \tilde{x}_q\bigr)\wedge\tilde{x}_m\wedge\rmd \tilde{x}_n\,,
\end{align}
and this is para-Hermitian but not para-K\"ahler unless $\pi^{mn}$ is constant. 
\end{exmp}

\subsection{Canonical generalized Lie derivative}

Let us consider diffeomorphisms generated by the standard Lie derivative
\begin{align}
 \Lie_V W^I \equiv V^J\,\partial_J W^I - W^J\,\partial_J V^I \,.
\end{align}
Under the Lie derivative, the covariant derivative $\nafla_I f$ of an arbitrary tensor $f$ transforms covariantly. 
However, unlike the standard geometry, the diffeomorphisms are restricted to a small subgroup due to the section condition. 
Moreover, if we keep the canonical form \eqref{eq:etaIJ} of $\eta_{IJ}$\,, the diffeomorphisms are reduced to the global $\OO(d,d)$ transformation \cite{1502.02428}\footnote{A particular $B$-field gauge transformation $B_{mn}\to B_{mn}+\partial_{[m}\lambda_{n]}$ ($\partial_{(m}\lambda_{n)}=0$) is also allowed \cite{1502.02428}.}
\begin{align}
 x^I \to x'^I = \Lambda^I{}_J\,x^J\qquad \Lambda\in \OO(d,d)\,,
\end{align}
which is a symmetry of the conventional DFT action. 
The usual diffeomorphisms on the physical space $\cF_+$ are rather contained in another local symmetry, which is known as the generalized diffeomorphism. 
When the para-complex structure is given by $K=K^{(0)}$, this symmetry is generated by the generalized Lie derivative
\begin{align}
 \gLie^{(0)}_V W^I \equiv V^J\,\nafla_J W^I - \bigl(\nafla_J V^I - \nafla^I V_J\bigr)\,W^J \,,
\end{align}
which never changes the flat metric: $\gLie_V\eta_{IJ}=0$\,. 
When the diffeomorphism parameter $V^I$ is a tangent vector field of the physical space $\cF_+$\,, i.e., $V^I=(\Pi_+V)^I = (v^m,\,0)$\,, the generalized Lie derivative reduces to the standard Lie derivative $\Lie_v$ on the physical space $\cF_+$,
\begin{align}
 \gLie^{(0)}_V W^I = \begin{pmatrix} \Lie_v w^m \\ \Lie_v \tilde{w}_m \end{pmatrix} ,\qquad 
 W^I \equiv \begin{pmatrix} w^m \\ \tilde{w}_m \end{pmatrix} .
\end{align}
In this sense, the generalized Lie derivative is a generalization of the usual Lie derivative. 
The para-complex structure $K^{(0)}$ is invariant under this transformation, but if the diffeomorphism parameter $V^I$ has a general form $(V^I)=(v^m,\,\tilde{v}_m)$, it is transformed as
\begin{align}
 \delta K^{(0)} = \gLie^{(0)}_V K^{(0)} = \begin{pmatrix} 0 & 0 \\ 4\,\partial_{[m}\tilde{v}_{n]} & 0 \end{pmatrix} .
\end{align}
Thus the diffeomorphism parameter $\tilde{v}_m$ causes a deformation of the foliation. 
In general, under a finite generalized diffeomorphism, $K^{(0)}$ is generally mapped to $K^{(b)}$ given in Eq.~\eqref{eq:K-b}. 

If we consider a general $K$, an issue arises for $\gLie^{(0)}_V$. 
For two vector fields $V=\Pi_+(V)$ and $W=\Pi_+(W)$ that are tangential to the physical space $\cF_+$\,, we have
\begin{align}
 \gLie^{(0)}_V W^I = \bigl[\Pi_+\bigl([V,\,W]\bigr)\bigr]^I - \cT_{J}{}^I{}_K\,V^J\,W^K \,,
\label{eq:gLie0}
\end{align}
where $\cT_{IJK}=\cT_{[IJK]}$ is defined as
\begin{align}
 \cT_{IJK} \equiv 3\,\Phi_{[IJK]}\,,\qquad 
 \Phi_I{}^K{}_J \equiv \frac{1}{2}\, K^K{}_L \,\nafla_I K^L{}_J \,.
\label{eq:def-cT-Phi}
\end{align}
Then, although both $V$ and $W$ are restricted to be tangent vectors on the physical space, the generalized Lie derivative does not reduce to the usual Lie derivative on the physical space due to the second term in Eq.~\eqref{eq:gLie0}. 
This prompts us to consider a modification of the generalized Lie derivative. 
In fact, there is the unique generalized Lie derivative that satisfies \cite{1806.05992}
\begin{align}
 \gLie_{\Pi_\pm(V)} \Pi_\pm(W) = \Pi_\pm\bigl([\Pi_\pm V,\,\Pi_\pm W]\bigr) \,.
\label{eq:gLie-canonical}
\end{align}
This is known as the canonical generalized Lie derivative and is defined by
\begin{align}
 \gLie_V W^I \equiv V^J\,\nabla_J W^I - \bigl(\nabla_J V^I - \nabla^I V_J\bigr)\,W^J \,.
\end{align}
Here, $\nabla$ is called the canonical connection\footnote{A short computation shows that canonical connection can be also expressed as
\begin{align*}
 \nabla_I V_J = (\Pi_+)_J{}^K\, \nafla_I (\Pi_+\,V)_K + (\Pi_-)_J{}^K\, \nafla_I (\Pi_-\,V)_K \,.
\end{align*}}
\begin{align}
 \nabla_I V_J \equiv \nafla_I V_J - \Phi_I{}^K{}_J\,V_K \,.
\end{align}
This is called para-Hermitian because it is compatible with the para-Hermitian structure,
\begin{align}
 \nabla_K \eta_{IJ} = 0\,,\qquad \nabla_K \omega_{IJ} = 0\,.
\end{align}
The difference between the two generalized Lie derivatives is called the generalized torsion
\begin{align}
 \cT(V,W) \equiv \cT_{IJ}{}^K\,V^I\,W^J\,\partial_K \equiv \gLie^{(0)}_V W - \gLie_V W \,, 
\end{align}
and the components $\cT_{IJ}{}^K$ can be expressed as in Eq.~\eqref{eq:def-cT-Phi}. 
Then the canonical generalized Lie derivative can be expressed also as
\begin{align}
 \gLie_V W^I \equiv V^J\,\partial_J W^I - \bigl(\partial_J V^I - \partial^I V_J\bigr)\,W^J + \cT_J{}^I{}_K \,V^J\,W^K \,,
\end{align}
and we can easily show the property \eqref{eq:gLie-canonical} by using Eq.~\eqref{eq:gLie0}. 
Accordingly, we consider that the gauge symmetry of DFT for a general foliation is generated by the canonical generalized Lie derivative. 
In the case of para-K\"ahler manifolds (where $\rmd \omega=0$), we can easily see that $\gLie_V$ reduces to the conventional one $\gLie^{(0)}_V$ by using the identity
\begin{align}
 \cT_{IJK} = \frac{1}{2} \,\bigl[ (\rmd \omega)^{(3,0)} + (\rmd \omega)^{(2,1)} - (\rmd \omega)^{(1,2)} - (\rmd \omega)^{(0,3)} \bigr]_{IJK} \,.
\end{align}
In this sense, $\gLie_V$ is a modest modification of the generalized Lie derivative. 

\subsection{Born geometry}

In DFT, the dynamical metric and the Kalb--Ramond $B$-field are packaged into the generalized metric $\cH_{IJ}$ which satisfies
\begin{align}
 \cH_{IJ} = \cH_{JI}\,,\qquad \cH^I{}_K\,\cH^K{}_J = \delta^I_J \,. 
\label{eq:cH-DFT}
\end{align}
By choosing a frame where $K=K^{(0)}$ is realized, we parameterize the generalized metric as\footnote{See \cite{1707.03713} for a more general parameterization.}
\begin{align}
 \cH_{IJ} = \begin{pmatrix} (g-B\,g^{-1}\,B)_{mn} & B_{mp}\,g^{pn} \\ -g^{mp}\,B_{pn} & g^{mn} \end{pmatrix},
\label{eq:cH-param}
\end{align}
and interpret $g_{mn}$ as the metric on the physical space. 

The second condition of Eq.~\eqref{eq:cH-DFT} shows that $J^I{}_J \equiv \cH^I{}_J$ is an additional almost para-complex structure on the doubled space. 
This is sometimes called the chiral structure because this matrix defines a chirality of the string (see section \ref{sec:DSM}). 
If this chiral structure satisfies
\begin{align}
 J^K{}_I\,J^L{}_J\,\omega_{KL} = - \omega_{IJ}\,,
\label{eq:Born-condition}
\end{align}
an almost para-Hermitian manifold is called the Born manifold. 
In this case, we have an almost complex structure $I^I{}_J \equiv \cH^{IK}\,\omega_{KJ} = (J\,K)^I{}_J$ satisfying $I^2=-\bm{1}$\,, and the pair $(I,J,K)$ is called an almost para-quaternionic structure. 
By substituting the parameterization \eqref{eq:cH-param} into the condition \eqref{eq:Born-condition}, we obtain $B_{mn}=0$\,, which looks very strong. 
In order to relax the requirement, one may assume that $K$ and $\omega$ have the form \cite{1707.00312,1802.08180,1806.05992,1910.09997}
\begin{align}
 K^{(B)} \equiv \begin{pmatrix} \delta^m_n & 0 \\ 2\,B_{mn} & -\delta_m^n \end{pmatrix} ,\qquad 
 \omega^{(B)} \equiv \eta\,K^{(B)} = \begin{pmatrix} 2\,B_{mn} & -\delta_m^n \\ \delta^m_n & 0 \end{pmatrix} ,
\end{align}
in the duality frame where the generalized metric has the form \eqref{eq:cH-param}. 
This allows us to satisfy Eq.~\eqref{eq:Born-condition} but the integrability $N_+ =0$ is broken when the $B$-field is not closed. 
Even when the integrability $N_+ =0$ is broken, as long as the integrability $N_- =0$ is satisfied, we can define the physical space as $\cF_-$ (which satisfies $L_-=T\cF_-$), and we may consistently formulate the gauged Born sigma model \cite{1910.09997}. 

In this paper, we do not require Eq.~\eqref{eq:Born-condition} and do not include the supergravity fields, such as the $B$-field, into the fundamental 2-form $\omega$\,. 
Then, the generalized metric does not describe the Born geometry, but since the action studied in the next subsection has the same form as that of the gauged Born sigma model, we shall use the nomenclature, the Born sigma model. 
As we explain in the next subsection, what we include into $\omega$ are the field strengths of the worldvolume gauge fields. 

\subsection{Born sigma model for string}
\label{sec:DSM}

Here we consider the string sigma model. 
We suppose that the doubled space is a para-K\"ahler manifold with the structure $(\eta,\,\omega^{(0)})$, and consider the action
\begin{align}
 S = \frac{1}{2}\int_\Sigma\Bigl[\frac{1}{2}\Exp{\lambda}\cH_{IJ}(x)\,DX^I\wedge *_{\gamma} DX^J - \frac{\mu_1}{2}\,\omega^{(F)}_{IJ}\,DX^I\wedge DX^J \Bigr] \,.
\label{eq:DSM}
\end{align}
Here, $\mu_1$ is the string charge (or tension), $\cH_{IJ}(x)$ is a generalized metric satisfying the linear section condition $ (\Pi_-^{(0)})^J{}_I\,\nafla_J =0$, and $DX^I$ is defined by
\begin{align}
 DX^I(\sigma) \equiv \rmd X^I(\sigma) -\cA^I(\sigma)\,,\qquad
 X^I(\sigma) \equiv \begin{pmatrix} x^m(\sigma) \\ \tilde{x}_m(\sigma) \end{pmatrix}.
\end{align}
The scalar field $\lambda(\sigma)$ is an auxiliary field that determines the tension, and the gauge fields $\cA^I(\sigma)$ are defined as $\cA^I(\sigma) \equiv (\Pi^{(0)}_-)^{In} \,C_n(\sigma) = (0,\,C_m)$\,. 
In the second term, we have defined
\begin{align}
 \omega^{(F)}_{IJ} \equiv \eta_{IK}\,(K^{(F)})^K{}_J \,, \qquad
 (K^{(F)})^I{}_J \equiv \begin{pmatrix} \delta^m_n & 0 \\ 2\,F_{mn} & -\delta_m^n \end{pmatrix}, \qquad
 F_{mn} \equiv \partial_m a_n - \partial_n a_m \,,
\end{align}
and the second term can be expanded as
\begin{align}
 \frac{\mu_1}{2}\,\omega^{(F)}_{IJ}\,DX^I\wedge DX^J = \mu_1\,\bigl(D\tilde{x}_m\wedge \rmd x^m + 2\,F_2\bigr) \qquad (F_2 \equiv \rmd a_1\,,\ a_1\equiv a_m\,\rmd x^m) \,.
\end{align}
We note that the covariant derivative $DX^I$ is invariant under the gauge transformation
\begin{align}
 X^I\to X^I + (\Pi^{(0)}_-)^{In}\,\chi_n\,,\qquad C_m\to C_m + \rmd\chi_n\,.
\label{eq:gauge-C}
\end{align}
We also note that $K^{(F)}$ describes the foliation of the $d$-dimensional space that the string lives in. 
This is different from the foliation characterized by $K^{(0)}$ and the deviation is characterized by the field strength $F_2(\sigma)$. 
If there are several strings propagating on the physical space, each string can live in a different $d$-dimensional space and each foliation is determined dynamically.\footnote{In the string case, the gauge field $A_1$ appears only in the boundary term and it is not dynamical, but in the case of higher-dimensional branes, such as the M5-brane, gauge fields can be dynamical.} 
Then, different strings observe the physical space from different angles. 

The equations of motion for $\lambda$ and the gauge fields $C_m$ give $\Exp{\lambda}=\mu_1$ (for $\mu_1>0$) and
\begin{align}
 (\Pi^{(0)}_+)^I{}_J\,\bigl( DX^J - \cH^J{}_K\, *_{\gamma} DX^K \bigr) 
 = (\Pi^{(0)}_+)^I{}_n\,\bigl( \rmd x^n - \cH^n{}_K\, *_{\gamma} DX^K \bigr) =0 \,.
\label{eq:SD-string+}
\end{align}
By taking the Hodge dual of this equation, we obtain
\begin{align}
 (\Pi^{(0)}_+)^I{}_J\,\cH^J{}_K\,\bigl( DX^K - \cH^K{}_L\, *_{\gamma} DX^L \bigr) 
 = (\Pi^{(0)}_+)^I{}_n\,\cH^{np}\,\bigl( D\tilde{x}_p - \cH_{pL}\, *_{\gamma} DX^L \bigr) =0 \,.
\end{align}
If $\cH^{mn}=g^{mn}$ is invertible, this is equivalent to
\begin{align}
 (\Pi^{(0)}_-)^I{}_J\,\bigl( DX^J - \cH^J{}_K\, *_{\gamma} DX^K\bigr) =0 \,,
\label{eq:SD-string-}
\end{align}
and then Eqs.~\eqref{eq:SD-string+} and \eqref{eq:SD-string-} give the self-duality relation (or the chirality condition)
\begin{align}
 DX^I = \cH^I{}_J\, *_{\gamma} DX^J\,.
\end{align}
According to this relation, $J^I{}_J=\cH^I{}_J$ is called the chirality operator. 

If we consider a flat background, the equations of motion give $\rmd \cA^I=0$ and we can fix the gauge symmetry \eqref{eq:gauge-C} as $\cA^I=0$\,. 
Then we obtain $DX^I=\rmd X^I$ and the on-shell value of the action becomes
\begin{align}
 S = \frac{\mu_1}{2}\int_\Sigma\Bigl[\frac{1}{2}\,\cH_{IJ}\,\rmd X^I\wedge *_{\gamma} \rmd X^J + \rmd x^m\wedge \rmd\tilde{x}_m \Bigr] \,,
\end{align}
where we have truncated $F_2$ for the sake of comparison. 
This is precisely Hull's action \cite{hep-th:0406102,hep-th:0605149}, and the topological term $\rmd x^m\wedge \rmd\tilde{x}_m$ (which comes from the $\omega$ term) plays an important role, for example, in the computation of the partition function \cite{hep-th:0701080}. 

\paragraph{Boundary condition and D-brane\\}

Under the equations of motion, a variation of the action becomes
\begin{align}
 \delta S \overset{\text{e.o.m.}}{\sim} \int_{\partial\Sigma} *_\gamma \bm{\theta} \,,\qquad
 *_\gamma \bm{\theta} \equiv \frac{\mu_1}{2}\,\bigl(\eta+\omega^{(F)}\bigr)_{IJ}\, DX^I\,\delta X^J \,.
\end{align}
In order to make the variational principle well-defined, we require the boundary condition
\begin{align}
 2\,n_a\,\epsilon^{ab} \,(*_\gamma \bm{\theta})_b\bigr\rvert_{\partial\Sigma} 
 = \mu_1\, n_a\,\epsilon^{ab}\, \bigl(\eta+\omega^{(F)}\bigr)_{In}\, D_b X^I\,\delta x^n\bigr\rvert_{\partial\Sigma} = 0\,,
\label{eq:BC-string}
\end{align}
where $n_a$ denotes the vector field normal to the boundary $\partial\Sigma$. 
By introducing a projection operator $(\pi_D)^m{}_n$ which has only the diagonal elements with values $1$ or $0$\,, we can impose the Dirichlet boundary condition as
\begin{align}
 (\pi_D)^m{}_n \,\delta x^n\bigr\rvert_{\partial\Sigma} = 0\,.
\end{align}
Then, Eq.~\eqref{eq:BC-string} requires the Neumann boundary condition for the other directions
\begin{align}
 n_a\,\epsilon^{ab} \,\bigl(\eta+\omega^{(F)}\bigr)_{In}\, D_b X^I\,(\pi_N)^n{}_p\bigr\rvert_{\partial\Sigma} 
 = n_a\,\epsilon^{ab} \,(D_b \tilde{x}_m - F_{mn}\, \partial_b x^n)\,(\pi_N)^m{}_p\bigr\rvert_{\partial\Sigma} = 0\,,
\end{align}
where $(\pi_N)^m{}_n\equiv \delta^m_n -(\pi_D)^m{}_n$\,. 
By using the equations of motion
\begin{align}
 D \tilde{x}_m = g_{mn}\,*_\gamma \rmd x^n +B_{mn}\, \rmd x^n \,,
\end{align}
the Neumann condition can be also expressed in the standard form
\begin{align}
 n_a\,\epsilon^{ab}\,\bigl[ g_{mn}\,\varepsilon^c{}_b \,\partial_c x^n + (B - F)_{mn}\, \partial_b x^n \bigr]\,(\pi_N)^m{}_p\bigr\rvert_{\partial\Sigma} = 0\,.
\end{align}
By noting that the Dirichlet boundary condition can be also written as
\begin{align}
 n_a\,\epsilon^{ab} \,(\pi_D)^m{}_n \,D_b x^n\bigr\rvert_{\partial\Sigma} = 0\,,
\label{eq:Dirichlet-string}
\end{align}
both the Dirichlet and the Neumann boundary conditions can be summarized as
\begin{align}
 n_a\,\epsilon^{ab} \,(\Pi^{(F)}_D)^I{}_J \,D_b X^J\bigr\rvert_{\partial\Sigma} = 0\,,
\end{align}
where
\begin{align}
 \Pi^{(F)}_D \equiv \Exp{\bm{F}}\Pi_D\Exp{-\bm{F}}\,,\qquad 
 \Pi_D \equiv \begin{pmatrix} (\pi_D)^m{}_n & 0 \\ 0 & (\pi_N)^n{}_m \end{pmatrix} ,\qquad
 \Exp{\bm{F}} \equiv \begin{pmatrix} \delta^m_n & 0 \\ F_{mn} & \delta_m^n \end{pmatrix}.
\end{align}

If we again consider a flat background, $DX^I$ can be gauge fixed to $\rmd X^I$ and the boundary condition reduces to
\begin{align}
 n_a\,\epsilon^{ab} \,(\Pi^{(F)}_D)^I{}_J \,\partial_b X^J\bigr\rvert_{\partial\Sigma} = 0\,.
\end{align}
This can be interpreted as a generalized Dirichlet boundary condition in the doubled space \cite{hep-th:0406102} that extends the conventional one \eqref{eq:Dirichlet-string}. 
Since there are $d$ ``$+1$'' in the diagonal elements of $\Pi^{(F)}_D$, regardless of the choice of the matrix $\pi_D$, the string is always attached to the ``generalized Dirichlet brane'' which is a $d$-dimensional object in the doubled space. 
In particular, when this object behaves as a $p$-brane in the physical space (namely when the trace of $\pi_N$ is $p+1$), this object is called a D$p$-brane \cite{hep-th:0406102}. 
In this way, the double sigma model (or the Born sigma model) allows us to describe the D$p$-brane with various values of $p$ as a single $d$-dimensional object in the doubled space \cite{hep-th:0406102}. 

\section{Exceptional space}
\label{sec:exceptional}

Here, we consider an extension of the same idea to the $U$-duality-covariant formulation, known as the exceptional field theory (EFT) \cite{hep-th:0104081,hep-th:0307098,1008.1763,1111.0459,1206.7045,1208.5884,1308.1673,1312.0614,1312.4542,1406.3348}. 
In the $E_n$ EFT, we introduce an exceptional space with local coordinates $x^I$ which transform in the $R_1$-representation of the $E_n$ $U$-duality group. 
In the exceptional space, the section condition can be expressed as\footnote{Additional conditions appear if we consider $n\geq 7$ but they do not affect the discussion here.}
\begin{align}
 \eta^{IJ;\cK}\,\partial_I f\, \partial_J g = 0\,.
\label{eq:SC-EFT}
\end{align}
Here, $\eta^{IJ;\cK}$ is an intertwining operator (called the $\eta$-symbol\footnote{In \cite{0906.1177} it is denoted as $f^{M_1N_1}{}_{P_2}$ but here we follow the notation of \cite{1610.01620,1708.06342}.}) which connects a symmetric product of the $R_1$-representation and the $R_2$-representation (labeled by $\cI,\cJ,\cK,\cdots$). 
An important difference from the DFT case \eqref{eq:SC} is that the $R_2$-representation is not a singlet.\footnote{Another difference is that the construction of the covariant derivative satisfying $\nafla_I \eta=0$ is non-trivial in EFT due to the last index of $\eta^{IJ;\cK}$ (see \cite{1705.09304} for a discussion on a such connection in the $\SL(5)$ EFT). Here we restrict ourselves to coordinate systems where $\eta^{IJ;\cK}$ is constant such that $\nafla_I$ is reduced to $\partial_I$\,.} 

In order to satisfy the section condition \eqref{eq:SC-EFT}, we again introduce a projector satisfying
\begin{align}
 \Pi_+^2 = \Pi_+\,,\qquad 
 \eta^{KL;\cI}\,(\Pi_+)^I{}_K\,(\Pi_+)^J{}_L = 0\,,
\label{eq:Piplus-EFT}
\end{align}
and assume that any fields $f$ satisfy the linear section condition
\begin{align}
 \partial_I f = (\Pi_+)^J{}_I\,\partial_J f\,. 
\end{align}
Solution of the section condition has been studied in the literature (see for example \cite{1308.1673,1311.5109}), and there are two inequivalent solutions: the M-theory section and the type IIB section. 
The rank of the projector is $n$ in the former case while $d\equiv n-1$ in the latter. 
For each case, we can introduce a projector $\Pi_-$ and $K$ as
\begin{align}
 (\Pi_-)^I{}_J \equiv \delta^I_J - (\Pi_+)^I{}_J\,,\qquad K^I{}_J \equiv (\Pi_+)^I{}_J - (\Pi_-)^I{}_J \,.
\end{align}
Unlike DFT, the dimensions of the $\pm1$ eigenspace of $K$ are different, and we call $K$ an almost product structure (rather than an almost complex structure). 
Moreover, we define
\begin{align}
 \omega_{IJ;\cK} \equiv \eta_{IL;\cK}\,K^L{}_J \,,
\end{align}
which naturally extends the fundamental 2-form $\omega_{IJ}$ in DFT [recall Eq.~\eqref{eq:omega-DFT}]. 
Similar to the DFT case, we assume $\eta_{IL;\cK}$ is constant but $K$ can be coordinate-dependent, and the integrability of $K$ is not ensured in general. 

In the following, we show the explicit form of the matrices $K$ and $\omega_{IL;\cK}$, and after that we discuss brane actions.
For convenience, we employ the following notation for multiple indices. 
For example, $A_{\bar{i}_p}$ represents $\frac{A_{[i_1\cdots i_p]}}{\sqrt{p!}}$\,. 
The factorial is introduced in order to reduce the overcounting: $A_{\bar{i}_p}\,B^{\bar{i}_p}=\frac{A_{[i_1\cdots i_p]}}{\sqrt{p!}}\,\frac{B^{[i_1\cdots i_p]}}{\sqrt{p!}}=\frac{1}{p!}\,A_{[i_1\cdots i_p]}\,B^{[i_1\cdots i_p]}$\,. 
The antisymmetrized Kronecker delta is defined as $\delta^{i_1\cdots i_p}_{j_1\cdots j_p}\equiv \delta^{[i_1}_{[j_1}\cdots \delta^{i_p]}_{j_p]}$ and we also define $\bm{\delta}^{i_1\cdots i_p}_{j_1\cdots j_p}\equiv p!\,\delta^{i_1\cdots i_p}_{j_1\cdots j_p}$\,. 
In the multiple-index notation, the latter Kronecker delta is denoted as $\delta^{\bar{i}_p}_{\bar{j}_p}\equiv \frac{\bm{\delta}^{i_1\cdots i_p}_{j_1\cdots j_p}}{\sqrt{p!}\sqrt{p!}}=\delta^{i_1\cdots i_p}_{j_1\cdots j_p}$\,, which satisfies $\delta^{\bar{i}_p}_{\bar{k}_p}\,\delta^{\bar{k}_p}_{\bar{j}_p} = \delta^{\bar{i}_p}_{\bar{j}_p}$\,. 
For example, for a 6-form $F_6$ and a $3$-form $F_3$\,, we have
\begin{align}
 F_{\bar{i}_5j} + \frac{1}{2!}\,\delta^{\bar{k}_3\bar{l}_2}_{\bar{i}_5}\,F_{\bar{k}_3}\,F_{\bar{l}_2j} 
 &=\frac{F_{i_1\cdots i_5 j}}{\sqrt{5!}} + \frac{1}{2!}\,\frac{\bm{\delta}^{k_1k_2l_1l_2l_3}_{i_1\cdots i_5}}{\sqrt{3!\,2!\,5!}} \,\frac{F_{k_1k_2k_3}}{\sqrt{3!}}\,\frac{F_{l_1l_2 j}}{\sqrt{2!}} 
\nn\\
 &= \frac{F_{i_1\cdots i_5 j}+5\,F_{[i_1i_2i_3}\,F_{i_4i_5] j}}{\sqrt{5!}}\,.
\end{align}
We also use a bracket notation, such as $A_{[\bar{i}_p}\,B_{\bar{i}_q} \cdots C_{\bar{i}_{r}]} \equiv \delta_{\bar{i}_{p+q+\cdots +r}}^{\bar{k}_p\bar{l}_q\cdots \bar{m}_r}\,A_{\bar{k}_p}\,B_{\bar{l}_q}\cdots C_{\bar{m}_r}$\,. 
In the standard notation, this corresponds to
\begin{align}
 A_{[\bar{i}_p}\,B_{\bar{i}_q} \cdots C_{\bar{i}_r]} = \frac{(p+q+\cdots+r)!}{p!\,q! \cdots r!}\,\frac{A_{[i_1\cdots i_p}\,B_{i_{p+1}\cdots i_{p+q}}\cdots C_{i_{p+q+\cdots+1}\cdots i_{p+q+\cdots+r}]}}{\sqrt{(p+q+\cdots+r)!}}\,.
\end{align}
It is noted that this bracket does not always coincide with the standard one even for single indices
\begin{align}
 A_{[\bar{i}}\,B_{\bar{j}]} &= \delta_{\bar{i}\bar{j}}^{\bar{k}\bar{l}}\,A_{\bar{k}}\,B_{\bar{l}} = \bm{\delta}_{ij}^{kl}\,A_{k}\,B_{l} = 2\,A_{[i}\,B_{j]}\,,
\\
 A_{[\bar{i}\bar{j}]} &= \delta_{\bar{i}\bar{j}}^{\bar{k}_2}\,A_{\bar{k}_2} = \frac{\bm{\delta}_{ij}^{k_1k_2}}{\sqrt{2!}}\,\frac{A_{k_1k_2}}{\sqrt{2!}} = A_{[ij]}\,,
\end{align}
where $\bar{i}$ denotes a single index $\bar{i}_1$\,. 
Accordingly, when the bracket is defined in the modified sense, we should keep the bar $\bar{i}$ inside the bracket. 
Let us give an another example.
For a $p$-form $A_p$ and a $(q+1)$-form $B_{q+1}$\,, we obtain
\begin{align}
 A_{[\bar{i}_p}\,B_{\bar{j}_q\bar{k}]} &\equiv \bm{\delta}_{\bar{i}_p\bar{j}_q\bar{k}}^{\bar{m}_p\bar{n}_{q+1}}\, A_{\bar{m}_p}\, B_{\bar{n}_{q+1}}
 = \frac{(p+q+1)!}{p!\,(q+1)!}\,\delta_{i_1\cdots i_pj_1\cdots j_qk}^{m_1\cdots m_p n_1\cdots n_{q+1}}\,\frac{A_{m_1\cdots m_p}\,B_{n_1\cdots n_{q+1}}}{\sqrt{p!}\,\sqrt{q!}}
\nn\\
 &= \frac{\frac{(p+q+1)!}{p!\,q!}\, A_{[i_1\cdots i_p}\,B_{j_1\cdots j_qk]}}{\sqrt{p!\,q!}} = \frac{(A\wedge B)_{i_1\cdots i_p j_1\cdots j_qk}}{\sqrt{p!\,q!}}\,.
\end{align}
These notations are useful to remove unimportant numerical factors from various expressions. 

\subsection{M-theory section}

When we consider the M-theory section, we expand the $R_1$-representation as
\begin{align}
 (x^I) = \bigl(x^i,\, y_{\bar{i}_2} ,\,y_{\bar{i}_5},\,\cdots \bigr) \,,
\end{align}
and the $R_2$-representation as
\begin{align}
 (\eta^{IJ;\cK}) = \bigl(\eta^{IJ;k},\, \eta^{IJ;\bar{k}_4} ,\, \cdots \bigr)\,.
\end{align}
We note that the dimensions of these representations are finite for $n\leq 8$, but they are infinite for $n=9,10,11$. 
In addition to $\eta^{IJ;\cK}$\,, we also introduce
\begin{align}
 (\eta_{IJ;\cK}) = \bigl(\eta_{IJ;k},\, \eta_{IJ;\bar{k}_4} ,\, \cdots \bigr)\,,
\end{align}
which has the same matrix form as $\eta^{IJ;\cK}$ although the position of the indices are upside down. 
The explicit forms of $\eta_{IJ;\cK}$ are as follows:
\begin{align}
 &\eta_k 
 \equiv \begin{pmatrix}
 0 & \delta_{\bar{i}\bar{k}}^{\bar{j}_2} & 0 \\
 \delta_{\bar{j}\bar{k}}^{\bar{i}_2} & 0 & 0 & \cdots \\
 0 & 0 & 0 \\
 & \vdots & & \ddots
 \end{pmatrix} , \qquad
 \eta_{\bar{k}_4} 
 \equiv \begin{pmatrix}
 0 & 0 & \delta_{\bar{i}\bar{k}_4}^{\bar{j}_5} \\
 0 & \delta_{\bar{k}_4}^{\bar{i}_2\bar{j}_2} & 0 & \cdots \\
 \delta_{\bar{j}\bar{k}_4}^{\bar{i}_5} & 0 & 0 \\
 & \vdots & & \ddots
 \end{pmatrix} , \quad \cdots \,.
\end{align}
Then, we can easily see that a matrix
\begin{align}
 \Pi^{(0)}_+ \equiv \begin{pmatrix} \delta^i_j & 0 & 0 \\ 0 & 0 & 0 & \cdots \\ 0 & 0 & 0 \\ &\vdots&& \ddots \end{pmatrix} ,
\end{align}
indeed satisfies the conditions \eqref{eq:Piplus-EFT}. 
Then, we obtain
\begin{align}
 \Pi^{(0)}_- = \begin{pmatrix}
 0 & 0 & 0 
\\
 0 & \delta_{\bar{i}_2}^{\bar{j}_2} & 0 & \cdots
\\
 0 & 0 & \delta_{\bar{i}_5}^{\bar{j}_5} \\ &\vdots&& \ddots 
 \end{pmatrix} ,\qquad 
 K^{(0)} = \begin{pmatrix}
 \delta^i_j & 0 & 0 
\\
 0 & -\delta_{\bar{i}_2}^{\bar{j}_2} & 0 & \cdots 
\\
 0 & 0 & -\delta_{\bar{i}_5}^{\bar{j}_5} \\ &\vdots&& \ddots 
 \end{pmatrix} , 
\end{align}
and the matrix forms of $(\omega_{IJ;\cK}) = \bigl(\omega_{IJ;k},\, \omega_{IJ;\bar{k}_4} ,\, \cdots \bigr)$ become
\begin{align}
 &\omega^{(0)}_k 
 = \begin{pmatrix}
 0 & - \delta_{\bar{i}\bar{k}}^{\bar{j}_2} & 0 \\
 \delta_{\bar{j}\bar{k}}^{\bar{i}_2} & 0 & 0 & \cdots \\
 0 & 0 & 0 \\ &\vdots&& \ddots 
 \end{pmatrix} , \qquad
 \omega^{(0)}_{\bar{k}_4} 
 = \begin{pmatrix}
 0 & 0 & - \delta_{\bar{i}\bar{k}_4}^{\bar{j}_5} \\
 0 & - \delta_{\bar{k}_4}^{\bar{i}_2\bar{j}_2} & 0 & \cdots \\
 \delta_{\bar{j}\bar{k}_4}^{\bar{i}_5} & 0 & 0 \\ &\vdots&& \ddots 
 \end{pmatrix} , \quad \cdots\,.
\end{align}
We note that, unlike the DFT case, they are not antisymmetric in general: only the matrix $\omega^{(0)}_k$ is. 
However, they play an important role in the brane actions, and we consider they are natural generalizations of the fundamental 2-form $\omega^{(0)}_{IJ}$ in DFT.\footnote{Under a decomposition $x^i=(x^m,x^z)$ where $m=1,\dotsc,n-1$ and $x^z$ denotes the coordinate on the M-theory circle, $\omega^{(0)}_z$ plays the role of $\omega^{(0)}_{IJ}$ in type IIA DFT.}

Similar to the DFT case, we can consider a more general $\Pi_\pm$ or $K$ by acting $U$-duality transformations (which generalize the $B$-transformation)
\begin{align}
 \Pi^{(F)}_+ &= {\footnotesize \begin{pmatrix}
 \delta^i_j & 0 & 0 
\\
 -F_{j\bar{i}_2} & 0 & 0 & \cdots
\\
 -(F_{j\bar{i}_5}+ \frac{1}{2!} F_{[\bar{i}_3}\,F_{\bar{i}_2]j}) & 0 & 0 \\ \vdots&&& \ddots 
 \end{pmatrix}} , \quad
 \Pi^{(F)}_- = {\footnotesize\begin{pmatrix}
 0 & 0 & 0 
\\
 F_{j\bar{i}_2} & \delta_{\bar{i}_2}^{\bar{j}_2} & 0 & \cdots
\\
 F_{j\bar{i}_5}+\frac{1}{2!} F_{[\bar{i}_3}\,F_{\bar{i}_2] j} & 0 & \delta_{\bar{i}_5}^{\bar{j}_5}\\ \vdots&&& \ddots 
 \end{pmatrix}} ,
\nn\\
 K^{(F)} &= {\footnotesize\begin{pmatrix}
 \delta^i_j & 0 & 0 
\\
 -2\,F_{j\bar{i}_2} & -\delta_{\bar{i}_2}^{\bar{j}_2} & 0 & \cdots
\\
 -2\,(F_{j\bar{i}_5}+\frac{1}{2!} F_{[\bar{i}_3}\,F_{\bar{i}_2] j}) 
 & 0 & -\delta_{\bar{i}_5}^{\bar{j}_5} \\ \vdots&&& \ddots 
 \end{pmatrix}} ,
\end{align}
where $F_3\equiv \rmd a_2$ and $F_6\equiv \rmd a_5$ are arbitrary closed 3- and 6-form fields. 
For these, $\omega$ become
\begin{align}
 \omega^{(F)}_k 
 &\equiv {\footnotesize\begin{pmatrix}
 2\,F_{ijk} & - \delta_{\bar{i}\bar{k}}^{\bar{j}_2} & 0 \\
 \delta_{\bar{j}\bar{k}}^{\bar{i}_2} & 0 & 0 & \cdots \\
 0 & 0 & 0 \\ \vdots&&& \ddots 
 \end{pmatrix}} , 
\quad
 \omega^{(F)}_{\bar{k}_4} 
 \equiv {\footnotesize\begin{pmatrix}
 2\,(F_{ij\bar{k}_4}-\frac{1}{2!}F_{[i \bar{k}_2} F_{\bar{k}_2]j}) & 0 & - \delta_{\bar{i}\bar{k}_4}^{\bar{j}_5} \\
 -2\,\delta^{\bar{i}_2}_{[\bar{k}_2}\,F_{\bar{k}_2]j} & - \delta_{\bar{k}_4}^{\bar{i}_2\bar{j}_2} & 0 & \cdots \\
 \delta_{\bar{j}\bar{k}_4}^{\bar{i}_5} & 0 & 0 \\ \vdots&&& \ddots 
 \end{pmatrix}} .
\end{align}

\subsection{Type IIB section}

When we consider the type IIB section, the $R_1$-representation is decomposed as
\begin{align}
 (x^I) = \bigl(x^m,\, y^\alpha_m ,\, y_{\bar{m}_3} ,\, y^\alpha_{\bar{m}_5}, \cdots\bigr) \,,
\label{eq:IIB-xI}
\end{align}
where $m,n,p=1,\dotsc,d\,(=n-1)$\,, $\alpha,\beta=1,2$ and $\bar{m}_p$ denotes the multiple index.
The $R_2$-representation is decomposed as
\begin{align}
 (\eta^{IJ;\cK}) &= \bigl(\eta^{IJ}_\alpha,\, \eta^{IJ;\bar{p}_2} ,\, \eta^{IJ;\bar{p}_4}_\alpha ,\, \cdots \bigr)\,,
\\
 (\eta_{IJ;\cK}) &= \bigl(\eta_{IJ}^\alpha,\, \eta_{IJ;\bar{p}_2} ,\, \eta_{IJ;\bar{p}_4}^\alpha ,\, \cdots \bigr)\,.
\end{align}
Their explicit forms are given as follows:
\begin{align}
 \eta^\gamma&\equiv \begin{pmatrix} 0 & \delta_\beta^\gamma\,\delta_{m}^{n} & 0 & 0 \\
 \delta_\alpha^\gamma\,\delta^{m}_{n} & 0 & 0 & 0 & \cdots \\
 0 & 0 & 0 & 0 \\
 0 & 0 & 0 & 0 \\ & \vdots & & & \ddots 
 \end{pmatrix}, \quad
 \eta_{\bar{p}_2} \equiv \begin{pmatrix}
 0 & 0 & \delta_{\bar{m}\bar{p}_2}^{\bar{n}_3} & 0 \\
 0 & \epsilon_{\alpha\beta}\,\delta_{\bar{p}_2}^{\bar{n}\bar{m}} & 0 & 0 & \cdots \\
 \delta_{\bar{n}\bar{p}_2}^{\bar{m}_3} & 0 & 0 & 0 \\
 0 & 0 & 0 & 0 \\ & \vdots & & & \ddots 
 \end{pmatrix},
\\
%%%%%
 \eta_{\bar{p}_4}^\gamma 
 &\equiv \begin{pmatrix}
 0 & 0 & 0 & \delta_{\beta}^{\gamma}\,\delta_{\bar{p}_4\bar{m}}^{\bar{n}_5} \\
 0 & 0 & \delta_{\alpha}^{\gamma}\,\delta_{\bar{p}_4}^{\bar{m}\bar{n}_3} & 0 & \cdots \\
 0 & \delta_{\beta}^{\gamma}\,\delta_{\bar{p}_4}^{\bar{n}\bar{m}_3} & 0 & 0 \\
 \delta_{\alpha}^{\gamma}\,\delta_{\bar{p}_4\bar{n}}^{\bar{m}_5} & 0 & 0 & 0 \\ & \vdots & & & \ddots 
 \end{pmatrix} ,
\quad \cdots \,.
\end{align}
Similar to the M-theory case, we can compute $\Pi_\pm^{(F)}$ and $K^{(F)}$ as follows:
\begin{align}
 \Pi_+^{(F)} &= {\footnotesize \begin{pmatrix}
 \delta_n^m & 0 & 0 & 0 \\
 -F_{nm}^\alpha & 0 & 0 & 0 & \cdots \\
 -(F_{n\bar{m}_3} + \frac{1}{2!}\,\epsilon_{\gamma\delta}\, F^\gamma_{n[\bar{m}_1}\, F^\delta_{\bar{m}_2]}) & 0 & 0 & 0 \\
 -(F^\alpha_{n\bar{m}_5} + F_{n[\bar{m}_3}\,F^\alpha_{\bar{m}_2]} + \frac{1}{3!}\,\epsilon_{\gamma\delta}\, F^\gamma_{n[\bar{m}_1}\,F^\delta_{\bar{m}_2}\,F^\alpha_{\bar{m}_2]}) & 0 & 0 & 0 \\ \vdots &&&& \ddots 
\end{pmatrix}},
\label{eq:IIB-Pi+}
\\
 \Pi_-^{(F)} &= {\footnotesize 
 \begin{pmatrix}
 0 & 0 & 0 & 0 \\
 F_{nm}^\alpha & \delta^\alpha_\beta\, \delta_m^n & 0 & 0 & \cdots \\
 F_{n\bar{m}_3} + \frac{1}{2!}\,\epsilon_{\gamma\delta}\, F^\gamma_{n[\bar{m}_1}\, F^\delta_{\bar{m}_2]} & 0 & \delta_{\bar{m}_3}^{\bar{n}_3} & 0 \\
 F^\alpha_{n\bar{m}_5} + F_{n[\bar{m}_3}\, F^\alpha_{\bar{m}_2]} + \frac{1}{3!}\,\epsilon_{\gamma\delta}\, F^\gamma_{n[\bar{m}_1}\, F^\delta_{\bar{m}_2}\, F^\alpha_{\bar{m}_2]} & 0 & 0 & \delta^\alpha_\beta\, \delta_{\bar{m}_5}^{\bar{n}_5} \\ \vdots &&&& \ddots 
\end{pmatrix}, }
\label{eq:IIB-Pi-}
\\
 K^{(F)} &= {\footnotesize 
 \begin{pmatrix}
 \delta_n^m & 0 & 0 & 0 \\
 -2\,F_{nm}^\alpha & -\delta^\alpha_\beta\, \delta_m^n & 0 & 0 & \cdots \\
 -2\,(F_{n\bar{m}_3} + \frac{1}{2!}\,\epsilon_{\gamma\delta}\, F^\gamma_{n[\bar{m}_1}\, F^\delta_{\bar{m}_2]}) & 0 & -\delta_{\bar{m}_3}^{\bar{n}_3} & 0 \\
 -2\,(F^\alpha_{n\bar{m}_5} + F_{n[\bar{m}_3}\, F^\alpha_{\bar{m}_2]} + \frac{1}{3!}\,\epsilon_{\gamma\delta}\, F^\gamma_{n[\bar{m}_1}\, F^\delta_{\bar{m}_2}\, F^\alpha_{\bar{m}_2]}) & 0 & 0 & -\delta^\alpha_\beta\, \delta_{\bar{m}_5}^{\bar{n}_5} \\ \vdots &&&& \ddots 
 \end{pmatrix},
 }
\end{align}
where $F_2^\alpha\equiv\rmd a_1^\alpha$, $F_4\equiv \rmd a_3$, and $F_6^\alpha\equiv \rmd a_5^\alpha$ are arbitrary closed forms. 
It is noted that the trace of the projector $(\Pi_+^{(F)})^I{}_I$ is $n$ in the M-theory section while $(n-1)$ in the type IIB section. 
Thus $\Pi_\pm^{(F)}$ or the almost product structures in the M-theory section and type IIB sections cannot be related by a $U$-duality transformation. 

By using the almost product structure $K^{(F)}$, we obtain the matrices $\omega^{(F)}$ as follows:
\begin{align}
 (\omega^{(F)})^\gamma&\equiv {\footnotesize\begin{pmatrix} 2\, F_{mn}^\gamma & -\delta_\beta^\gamma\,\delta_{m}^{n} & 0 & 0 \\
 \delta_\alpha^\gamma\,\delta^{m}_{n} & 0 & 0 & 0 & \cdots \\
 0 & 0 & 0 & 0 \\
 0 & 0 & 0 & 0 \\ &\vdots &&& \ddots 
 \end{pmatrix} },
\\
%%%%%
 (\omega^{(F)})_{\bar{p}_2}&\equiv {\footnotesize \begin{pmatrix}
 2\,(F_{mn\bar{p}_2} - \frac{1}{2!}\,\epsilon_{\gamma\delta}\, F^\gamma_{n[\bar{m}}\, F^\delta_{\bar{p}_2]}) & 0 & - \delta_{\bar{m}\bar{p}_2}^{\bar{n}_3} & 0 \\
 -2\,\epsilon_{\alpha\beta}\,\delta_{[\bar{p}_1}^{\bar{m}}\,F_{\bar{p}_1] n}^\beta & - \epsilon_{\alpha\beta}\,\delta_{\bar{p}_2}^{\bar{n}\bar{m}} & 0 & 0 & \cdots \\
 \delta_{\bar{n}\bar{p}_2}^{\bar{m}_3} & 0 & 0 & 0 \\
 0 & 0 & 0 & 0 \\ \vdots &&&& \ddots 
 \end{pmatrix}},
\\
%%%%%
 (\omega^{(F)})_{\bar{p}_4}^\gamma 
 &\equiv {\footnotesize {\arraycolsep=0mm \begin{pmatrix}
 2\,(F^\gamma_{mn\bar{p}_4} - F_{n[\bar{p}_3}\,F^\gamma_{\bar{p}_1\bar{m}]} - \frac{1}{3!}\,\epsilon_{\alpha\beta}\, F^\alpha_{n[\bar{p}_1}\, F^\beta_{\bar{p}_2}\,F^\gamma_{\bar{p}_1\bar{m}]}) & 0 & 0 & - \delta_{\beta}^{\gamma}\,\delta_{\bar{p}_4\bar{m}}^{\bar{n}_5} \\
 2 \,\delta^\gamma_\alpha\,(F_{n[\bar{p}_3} + \frac{1}{2!}\,\epsilon_{\gamma\delta}\, F^\gamma_{n[\bar{p}_1}\, F^\delta_{\bar{p}_2})\,\delta^m_{\bar{p}_1]} & 0 & - \delta_{\alpha}^{\gamma}\,\delta_{\bar{p}_4}^{\bar{m}\bar{n}_3} & 0 & \cdots \\
 -2\, F^\gamma_{n[\bar{p}_1}\,\delta^{\bar{m}_3}_{\bar{p}_3]} & - \delta_{\beta}^{\gamma}\,\delta_{\bar{p}_4}^{\bar{n}\bar{m}_3} & 0 & 0 \\
 \delta_{\alpha}^{\gamma}\,\delta_{\bar{p}_4\bar{n}}^{\bar{m}_5} & 0 & 0 & 0 \\ \vdots &&&& \ddots 
 \end{pmatrix}}} .
\end{align}

\subsection{Generalized Lie derivative in EFT}
\label{sec:Lie-EFT}

In the conventional formulation of EFT, the generalized Lie derivative is defined as \cite{1208.5884}
\begin{align}
 \gLie_V^{(0)} W^I \equiv V^J\,\partial_J W^I - \partial_J V^I + Y^{IK}_{LJ}\,\partial_K V^L \,W^J\,,
\end{align}
where $Y^{IJ}_{KL}$ is an invariant tensor (e.g., $Y^{IJ}_{KL} = \eta^{IJ;\cI}\,\eta_{KL;\cI}$ for $n\leq 6$). 
However, similar to the DFT case, if we consider a non-constant polarization $\Pi_\pm$\,, this may not satisfy the property
\begin{align}
 \gLie^{(0)}_{\Pi_+(V)} \Pi_+(W) = \Pi_+\bigl[\Pi_+(V),\, \Pi_+(W)\bigr] \,,
\end{align}
meaning that the $\gLie^{(0)}$ does not reduce to the standard Lie derivative on the physical subspace. 
This issue can be resolved by considering a modification of the generalized Lie derivative similar to the DFT case. 
We redefine the generalized Lie derivative as
\begin{align}
 \gLie_V W^I \equiv V^J\,\nabla_J W^I - \bigl(\nabla_J V^I - Y^{IK}_{LJ}\,\nabla_K V^L \bigr)\,W^J\,,
\end{align}
where $\nabla_I V^J$ is defined as
\begin{align}
 \nabla_I V^J \equiv \partial_I V^J + \Phi_I{}^J{}_K\,V^K\,,\qquad 
 \Phi_I{}^J{}_K \equiv \frac{1}{2}\, K^J{}_L \,\partial_I K^L{}_K\,,
\end{align}
satisfying
\begin{align}
 \nabla_I K^J{}_K = 0 \,,\qquad \nabla_I (\Pi_\pm)^J{}_K = 0\,.
\end{align}
Similar to the DFT case, defining the generalized torsion as
\begin{align}
 \cT_{I}{}^{J}{}_{K} \equiv 2\,\Phi_{[I}{}^{J}{}_{K]} + Y^{JP}_{QK}\,\Phi_P{}^{Q}{}_{I}\,, 
\end{align}
we obtain
\begin{align}
 \gLie_V W^I \equiv V^J\,\partial_J W^I -W^J\,\partial_J V^I + Y^{IK}_{LJ}\,\partial_K V^L \,W^J + \cT_J{}^I{}_K\,V^J\,W^K \,.
\end{align}
We note that this kind of modified generalized Lie derivative has been studied in several contexts \cite{1705.09304,1708.02589}. 

If we require that tangent vectors on the physical subspace are maximally isotropic, i.e.,
\begin{align}
 Y^{IJ}_{KL}\,(\Pi_+)^K{}_P\,(\Pi_+)^L{}_Q = 0\,,
\end{align}
we can show
\begin{align}
 \gLie_{\Pi_+(V)} \Pi_+(W) = \Pi_+\bigl[\Pi_+(V),\, \Pi_+(W)\bigr] \,,
\end{align}
similar to the DFT case. 
However, in the exceptional space, according to the existence of many dual directions, we cannot require $Y^{IJ}_{KL}\,(\Pi_-)^K{}_P\,(\Pi_-)^L{}_Q = 0$\,, and we do not have the property $\gLie_{\Pi_-(V)} \Pi_-(W) = \Pi_-\bigl[\Pi_-(V),\, \Pi_-(W)\bigr]$ unlike the DFT case. 
For the integrability of the physical space, at least, we require
\begin{align}
 \Pi_-\bigl[\Pi_+(V),\, \Pi_+(W)\bigr] = 0 \quad
 \Leftrightarrow\quad 2\,\Phi_{[J}{}^I{}_{K]}\, (\Pi_+)^J{}_P\, (\Pi_+)^K{}_Q =0\,,
\end{align}
which gives a constraint for the generalized torsion. 
For the consistent formulation of EFT, we may need additional conditions, but here we do not study in further details. 
Of course, when the generalized torsion vanishes, the conventional EFT is recovered. 

In the following, we study brane actions by using two product structures $K^{(0)}$ and $K^{(F)}$ with vanishing generalized torsions $\cT_I{}^J{}_K$\,. 
The former defines the physical subspace and the supergravity fields satisfy $(\Pi^{(0)}_-)^J{}_I\,\partial_J =0$\,. 
On the other hand, the latter describes the foliation associated with the brane. 
This is described by the closed forms, collectively denoted by $F$\,. 
They correspond to the field strengths of the standard worldvolume gauge fields and in general dynamical. 

\subsection{Brane actions in M-theory}

We consider the action
\begin{align}
 S = \frac{1}{p+1}\int_\Sigma \Bigl[\Exp{\lambda}\frac{1}{2}\,\cH_{IJ}(X)\,DX^I\wedge *_{\gamma} DX^J -\frac{1}{2}\,\omega^{(F)}_{IJ;\cK}\, DX^I\wedge DX^J\wedge q^\cK_{(\text{brane})}\Bigr] \,,
\label{eq:M-action}
\end{align}
which is a natural extension of Eq.~\eqref{eq:DSM}. 
Here, $\cH_{IJ}(X)$ is a generalized metric (see Appendix \ref{app:conventions-M} for our conventions) and $DX^I$ is defined as
\begin{align}
 DX^I \equiv \rmd X^I(\sigma) - \cA^I(\sigma)\,,\qquad 
 \cA^I(\sigma)\equiv (\Pi^{(0)}_-)^{I\bar{J}} \,C_{\bar{J}}(\sigma)\,,
\end{align}
where $\bar{J}$ runs over the $R_1$-representation other than the physical directions denoted by $\{i\}$\,. 
Namely, we have
\begin{align}
 (DX^I) = \begin{pmatrix} \rmd x^i \\ \rmd y_{\bar{i}_2} - C_{\bar{i}_2} \\ \rmd y_{\bar{i}_5} - C_{\bar{i}_5} \\ \vdots\end{pmatrix} .
\end{align}
The $(p-1)$-form $q^\cK_{(\text{brane})}$ represents the charge vector associated with each brane. 
In this section, we consider the M2/M5-brane, and their corresponding charge vectors are given by (see Appendix \ref{app:charge-vector})
\begin{align}
 q^\cK_{(\text{M2})} \equiv \frac{\mu_2}{2}\begin{pmatrix} \rmd x^k \\ 0 \\ 0 \\ \vdots \end{pmatrix},\qquad
 q^\cK_{(\text{M5})} \equiv \frac{\mu_5}{5}\begin{pmatrix} -\rmd x^k\wedge F_3 \\ \rmd x^{\bar{k}_4} \\ 0 \\ \vdots \end{pmatrix},
\end{align}
where $\mu_p$ is a brane charge (or tension) and we have defined
\begin{align}
 \rmd x^{\bar{i}_p} \equiv \frac{\rmd x^{i_1}\wedge\cdots \wedge\rmd x^{i_p}}{\sqrt{p!}}\,. 
\end{align}
Note that, in the doubled space, the $R_2$-representation is a singlet and the charge vector is just a constant $q^\cK_{(\text{string})}=\mu_1$\,. 
Under this identification, the action \eqref{eq:M-action} reproduces \eqref{eq:DSM}. 

The second term in the action \eqref{eq:M-action} can be expanded as
\begin{align}
 \frac{1}{2}\,\omega^{(F)}_{IJ;\cK}\,DX^I \wedge DX^J\wedge q^\cK_{(\text{M2})}
 &= \mu_2\,\bigl(Dy_{\bar{i}_2}\wedge \rmd x^{\bar{i}_2} + 3\,F_3\bigr)\,, 
\label{eq:omega-M2}
\\
 \frac{1}{2}\,\omega^{(F)}_{IJ;\cK}\,DX^I \wedge DX^J\wedge q^\cK_{(\text{M5})}
 &= \mu_5\,\Bigl(Dy_{\bar{i}_5}\wedge \rmd x^{\bar{i}_5} - Dy_{\bar{i}_2}\wedge \rmd x^{\bar{i}_2}\wedge F_3 + 6\, F_6 \Bigr) \,.
\end{align}
Then, as naturally expected from the invariance of the action \eqref{eq:M-action} under the generalized Lie derivative, these actions are the same as the ones proposed in \cite{1607.04265,1712.10316} (note that the first term of Eq.~\eqref{eq:omega-M2} corresponds to the topological term proposed in Eq.~(9.1) of \cite{1802.00442}). 
As was shown there, they are (classically) equivalent to the standard (bosonic) M2/M5-brane theories. 

In \cite{1607.04265,1712.10316}, the discussion was restricted to $n\leq 7$\,, but such a restriction is not necessary. 
If we consider $n\geq 8$\,, the matrix size of the generalized metric $\cH_{IJ}$ becomes bigger and it can be infinite dimensional. 
However, the number of the auxiliary fields also increases accordingly.
Since the actions for the irrelevant auxiliary fields are always given by algebraic quadratic forms, after eliminating these, we obtain the brane actions that have the same as the one studied in $n\leq 7$.
The only difference is the range of the index $i=1,\dotsc,n$\,, and by choosing $n=11$\,, the full (bosonic) M2/M5-brane worldvolume theory in the 11D spacetime is recovered. 

\subsection{Brane actions in type IIB theory}

We can consider the same action also in type IIB theory
\begin{align}
 S = \frac{1}{p+1}\int_\Sigma \Bigl[\Exp{\lambda}\frac{1}{2}\,\cH_{IJ}(X)\,DX^I\wedge *_{\gamma} DX^J -\frac{1}{2}\,\omega^{(F)}_{IJ;\cK}\, DX^I\wedge DX^J\wedge q^\cK_{(\text{brane})}\Bigr]\,.
\label{eq:IIB-action}
\end{align}
Here, the $R_1$-representation is decomposed as in Eq.~\eqref{eq:IIB-xI}, and we choose the polarization tensor $\Pi^{(0)}_\pm$ as given in Eqs.~\eqref{eq:IIB-Pi+} and \eqref{eq:IIB-Pi-}. 
The parameterization of the generalized metric $\cH_{IJ}(X)$ is given in Appendix \ref{app:conventions-B}. 
The charge vectors associated with a $(p,q)$ string, D3-brane, and a $(p,q)$ 5-brane are respectively given by (see Appendix \ref{app:charge-vector})
\begin{align}
 q^\cK_{(p,q)\text{-1}} \equiv \mu_1{\footnotesize\begin{pmatrix} q_\gamma \\ 0 \\ 0 \\ 0 \\ \vdots \end{pmatrix}},\quad
 q^\cK_{(\text{D3})} \equiv \frac{\mu_3}{3}{\footnotesize\begin{pmatrix} -\epsilon_{\gamma\delta}\,F_2^\delta \\ \rmd x^{\bar{p}_2} \\ 0 \\ 0 \\ \vdots \end{pmatrix}},
\quad
 q^\cK_{(p,q)\text{-5}} \equiv \frac{\mu_5}{5} {\footnotesize\begin{pmatrix} q_\gamma\,F_4 + \frac{1}{2}\,q_\alpha\,\epsilon_{\gamma\beta}\,F^\alpha_2\wedge F_2^\beta \\ -q_\alpha\,\rmd x^{\bar{p}_2}\wedge F_2^\alpha \\ q_\gamma\,\rmd x^{\bar{p}_4} \\ 0 \\ \vdots \end{pmatrix}}.
\end{align}
We can again expand the second term in the action as
\begin{align}
 \frac{1}{2}\,\omega^{(F)}_{IJ;\cK}\,DX^I \wedge DX^J\wedge q^\cK_{(p,q)\text{-1}}
 &= \mu_1\,q_\alpha\,\bigl(Dy^\alpha_m\wedge \rmd x^m + 2\, F_2^\alpha \bigr)\,,
\\
 \frac{1}{2}\,\omega^{(F)}_{IJ;\cK}\,DX^I \wedge DX^J\wedge q^\cK_{(\text{D3})} 
 &= \mu_3\,\bigl(Dy_{\bar{m}_3}\wedge \rmd x^{\bar{m}_3} - \epsilon_{\alpha\beta}\, Dy_m^\alpha\wedge \rmd x^m\wedge F_2^\beta + 4\,F_4\bigr)\,, 
\\
 \frac{1}{2}\,\omega^{(F)}_{IJ;\cK}\,DX^I \wedge DX^J\wedge q^\cK_{(p,q)\text{-5}}
 &= \mu_5\,q_\alpha\,\bigl(Dy^\alpha_{\bar{m}_5}\wedge \rmd x^{\bar{m}_5} - Dy_{\bar{m}_3}\wedge \rmd x^{\bar{m}_3}\wedge F_2^\alpha + Dy_{m}^\alpha\wedge \rmd x^{m} \wedge F_4
\nn\\
 &\quad\ + \tfrac{1}{2}\, \epsilon_{\gamma\delta}\,Dy^\gamma_m\wedge\rmd x^m\wedge F^\delta_2\wedge F^\alpha_2
 + 2\,F_4\wedge F_2^\alpha + 6\,F_6^\alpha \bigr)\,.
\end{align}

As was shown in Ref.~\cite{1712.10316}, the string action reproduces the conventional one for the $(p,q)$-string (for a string, a similar $U$-duality-covariant sigma model is also discussed in \cite{1712.07115 ,1802.00442}). 
The actions for the D3-brane and the $(p,q)$ 5-branes have not been studied there. 
By eliminating the auxiliary fields, we find that these actions reproduce the following Wess--Zumino terms:
\begin{align}
 S_{\text{WZ}}^{(\text{D3})} &= \mu_3\int_\Sigma \bigl(A_4 -\tfrac{1}{2}\,\epsilon_{\alpha\beta}\,A_2^\alpha\wedge F_2^\beta - F_4\bigr)\,,
\\
 S_{\text{WZ}}^{(p,q)\text{-5}} &= \mu_5\,q_\alpha\int_\Sigma \bigl(A_6^\alpha -\tfrac{1}{3}\,A_4\wedge A_2^\alpha 
 -\tfrac{2}{3}\,A_4\wedge F_2^\alpha + \tfrac{1}{3}\,A_2^\alpha\wedge F_4
\nn\\
 &\qquad\qquad\quad -\tfrac{1}{6}\,\epsilon_{\gamma\delta}\,A_2^\gamma \wedge F_2^\delta\wedge F_2^\alpha
 - F_6^\alpha - \tfrac{1}{3}\,F_4^\alpha\wedge F_2^\alpha \bigr)\,. 
\end{align}
Apparently, they do not have the standard form. 
Indeed, the D3-brane action contains a doublet of the worldvolume gauge field strengths $F_2^\alpha$\,, although in the standard formulation we introduce only one gauge field. 
However, this kind of Wess--Zumino term that contains a doublet has been studied in $S$-duality-covariant formulations \cite{hep-th:9706208,hep-th:9708011,hep-th:9710007,hep-th:9804157,hep-th:9911201,hep-th:0611036}.
Then, it will be possible that the proposed theory is equivalent to the standard one after imposing a certain duality relation to the doublet of the gauge fields.\footnote{As was studied in detail in Ref.~\cite{1712.10316}, in the case of the M5-brane the equations of motion for the gauge field imply a self-duality relation, and a similar relation may be obtained also in the D3-brane case.} 
We will leave the consistency check with the standard formulation for future work. 

\subsection{Boundary conditions}

Unlike the doubled case, the boundary condition in the exceptional space is non-trivial. 

Before discussing higher-dimensional objects, let us consider the case of the $(p,q)$-string in type IIB theory (see \cite{1904.06714} for a related study), where a variation of the action becomes
\begin{align}
 \delta S \overset{\text{e.o.m.}}{\sim} \frac{1}{4}\int_{\partial\Sigma} \bigl(\eta+\omega^{(F)}\bigr)_{IJ;\cK}\, DX^I \,\delta X^J\,q^\cK_{(p,q)\text{-1}} 
 = \frac{q_\alpha}{2}\int_{\partial\Sigma} \bigl(Dy^\alpha_m - F^\alpha_{mn}\,\rmd x^n \bigr)\, \delta x^m \,.
\end{align}
By imposing the Dirichlet boundary condition [see Eq.~\eqref{eq:Dirichlet-string}]
\begin{align}
 (\pi_D)^m{}_n \,\delta x^n\bigr\rvert_{\partial\Sigma} = 0\quad \Leftrightarrow\quad 
 (\pi_D)^m{}_n\,n_a\,\epsilon^{ab} \,\partial_b x^n\bigr\rvert_{\partial\Sigma} = 0\,,
\end{align}
the Neumann boundary condition becomes
\begin{align}
 q_\alpha \, n_a\,\epsilon^{ab}\, \bigl(D_by^\alpha_m - F^\alpha_{mn}\,\partial_b x^n \bigr)\, (\pi_N)^m{}_p\bigr\rvert_{\partial\Sigma} = 0\,,
\end{align}
where $(\pi_N)^m{}_n\equiv \delta^m_n -(\pi_D)^m{}_n$\,.
The equations of motion for the auxiliary fields $\lambda$ and $\cA^I$ give
\begin{align}
 \Exp{\lambda} &= \abs{q}\,,\qquad \abs{q}\equiv \sqrt{q_\alpha\,m^{\alpha\beta}\,q_\beta}\,,
\\
 DX^I &= (L^{-1})^I{}_J\,\hat{P}^J\,,\qquad 
 (\hat{P}^I) \equiv {\footnotesize\begin{pmatrix} \rmd x^m \\ \frac{m^{\alpha\beta}\,q_\beta}{\abs{q}}\,\sfg_{mn}*_\gamma \rmd x^n \\ 0 \\ \vdots \end{pmatrix}},
\end{align}
where $m_{\alpha\beta}$ and $L^I{}_J$ are matrices including only the supergravity fields (see Appendix \ref{app:conventions-B}) and $\sfg_{mn}$ denotes the Einstein-frame metric. 
The action is then reduced to
\begin{align}
 S = \frac{\mu_1}{2}\int_\Sigma \abs{q}\,\sfg_{mn}\,\rmd x^m\wedge *_\gamma \rmd x^n + \mu_1\,q_\alpha \int_\Sigma \bigl(A^\alpha_2 - F_2^\alpha\bigr)\,. 
\end{align}
For simplicity, if we consider a flat background with $L^I{}_J=\delta^I_J$ (i.e., with vanishing $p$-form potentials), the equations of motion for $x^m(\sigma)$ lead to $\rmd\hat{P}^I=0$ and $DX^I$ is a closed form. 
Then we can realize $DX^I = \rmd X^I$\,, and the Neumann boundary condition becomes 
\begin{align}
 q_\alpha \, n_a\,\epsilon^{ab}\, \bigl(\partial_b y^\alpha_m - F^\alpha_{mn}\,\partial_b x^n \bigr)\, (\pi_N)^m{}_p\bigr\rvert_{\partial\Sigma} = 0\,.
\end{align}
Under this situation, we can combine the Dirichlet/Neumann boundary conditions as
\begin{align}
 &(\Pi^{(F)}_D)^I{}_J\,n_a\,\epsilon^{ab} \,\partial_b X^J\bigr\rvert_{\partial\Sigma} = 0\,,\qquad \Pi_D^{(F)} \equiv \Exp{\bm{F}}\Pi_D\,\Exp{-\bm{F}}\,,
\\
 &\Pi_D \equiv \begin{pmatrix}
 (\pi_D)^m{}_n & 0 & * \\
 0 & \frac{m^{\alpha\epsilon}\,q_\epsilon\,q_\beta}{q_\gamma\,m^{\gamma\delta}\,q_\delta}\,(\pi_N)^n{}_m & * & \cdots \\
 * & * & * \\
 & \vdots & & \ddots
\end{pmatrix},\qquad 
 \Exp{\bm{F}} \equiv 
 \begin{pmatrix}
 \delta^m_n & 0 & \cdots \\
 F^\delta_{mn} & \delta^\alpha_\beta\,\delta_m^n \\
 \vdots & & \ddots
\end{pmatrix}\,,
\end{align}
where $\Pi_D$ is a projection operator and $\Exp{\bm{F}}$ is an element of the $U$-duality group. 
Unfortunately, some elements of the Dirichlet projector $\Pi_D$ (denoted by ``$*$'') cannot be determined because the third or lower components of the generalized vector $\rmd X^I$ identically vanish
\begin{align}
 \rmd X^I = {\footnotesize\begin{pmatrix} \rmd x^m \\ \frac{m^{\alpha\beta}\,q_\beta}{\abs{q}}\,\sfg_{mn}*_\gamma \rmd x^n \\ 0 \\ \vdots \end{pmatrix}},
\end{align}
under the equations of motion. 
The Dirichlet projector $\Pi_D$ is defined to have the diagonal elements with values $1$ or $0$\,, and the number of the element $1$ [namely the trace $(\Pi_D)^I{}_I$] corresponds to the number of the Dirichlet directions in the exceptional space. 
The number of Dirichlet directions $d_D$ should correspond to the co-dimension of the extended object in the exceptional space. 
In the special cases where $n\leq 3$\,, the undetermined components disappear and we obtain the number of the Dirichlet directions as $d_D = (\pi_D)^m{}_m + (\pi_N)^m{}_m = n-1$\,. 
For $n\geq 4$\,, our analysis only gives the lower bound $d_D \geq n-1$\,.
According to the analysis based on the supersymmetry, it is claimed that $d_D = 2^{n-2}$ for $n\leq 7$ \cite{1904.06714}. 
This indicates that there exists an object with the co-dimension $2^{n-2}$\,, and it is interesting to study the effective theory of such a higher-dimensional object in the exceptional space. 

Instead of a string, we can also consider a higher-dimensional object, where we face a difficulty. 
As an example, let us consider a membrane action in M-theory. 
Under the equations of motion, a variation of the action becomes
\begin{align}
 \delta S \overset{\text{e.o.m.}}{\sim} \int_{\partial\Sigma} *_\gamma \bm{\theta} \,,
\end{align}
where a 1-form $\bm{\theta}$ is given by
\begin{align}
 *_\gamma \bm{\theta} &\equiv \frac{\mu_2}{8}\,\bigl[\bigl(\eta+\omega^{(F)}\bigr)_{IJ;k}\, DX^I\,\delta X^J\wedge \rmd x^k - \omega^{(F)}_{IJ;k}\, DX^I\wedge DX^J\, \delta x^k\bigr]
\nn\\
 &= - \frac{\mu_2}{2} \,\bigl(Dy_{ik} \wedge \rmd x^i + F_{ijk}\,\rmd x^i\wedge\rmd x^j \bigr)\, \delta x^k \,.
\end{align}
Then, the Dirichlet boundary condition is
\begin{align}
 (\pi_D)^i{}_j \,\delta x^j\bigr\rvert_{\partial\Sigma} = 0\,,\quad\Leftrightarrow\quad 
 (\pi_D)^i{}_j\, n_a\,\epsilon^{abc}\, \partial_b x^j \bigr\rvert_{\partial\Sigma} = 0\,,
\end{align}
while the Neumann boundary condition is
\begin{align}
 n_a\,\epsilon^{abc}\,\bigl(D_by_{ik}\,\partial_c x^i + F_{ijk}\,\partial_b x^i\partial_c x^j \bigr) \,(\pi_N)^k{}_l\bigr\rvert_{\partial\Sigma} = 0\,.
\end{align}
Unfortunately, $D_by_{ik}$ appears with the combination $D_by_{ik}\,\partial_c x^i$ in the Neumann boundary condition.
Consequently, it seems to be difficult to combine the Dirichlet and the Neumann boundary conditions as a single boundary condition
\begin{align}
 (\Pi_D)^I{}_J\, n_a\,\epsilon^{abc}\, D_b X^J \bigr\rvert_{\partial\Sigma} = 0\,.
\end{align}
Moreover, even if we assume that the target space is flat, $D X^I \equiv \rmd X^I -\cA^I$ is not a closed form under the equations of motion, and $\cA^I$ cannot be removed unlike the string case. 
Thus, it is difficult to discuss an extended object in the exceptional space where the membrane can end. 
The same difficulty exists also for other higher-dimensional objects where the charge vector $q^\cK_{\text{(brane)}}$ is not constant. 

\section{Brane actions in Hamiltonian form}
\label{sec:Hamiltonian}

In this section, we present brane actions in Hamiltonian form and see that the almost product structure $K$ again plays an important role there. 
For this purpose, we decompose the worldsheet coordinates into the temporal and the spatial directions as $(x^a)=(\tau,x^{\bar{a}})$ and decompose the intrinsic metric as
\begin{align}
 (\gamma_{ab}) = \begin{pmatrix} -N^2 + \sfh_{\bar{c}\bar{d}}\, N^{\bar{c}}\, N^{\bar{d}} & N^{\bar{c}}\,\sfh_{\bar{c}\bar{b}} \\ \sfh_{\bar{a}\bar{c}}\,N^{\bar{c}} & \sfh_{\bar{a}\bar{b}} \end{pmatrix},\qquad
 (\gamma^{ab}) = \begin{pmatrix} -\frac{1}{N^2} & \frac{N^{\bar{b}}}{N^2} \\ \frac{N^{\bar{c}}}{N^2} & \sfh^{\bar{a}\bar{b}}-\frac{N^{\bar{a}}\,N^{\bar{b}}}{N^2} \end{pmatrix}.
\end{align}

\subsection{String action}

Let us consider the first-order string action \cite{hep-th:0106042}
\begin{align}
 S_{1\text{st}} = \mu_1\int_\Sigma\Bigl(\rmd x^m\wedge \bm{P}_m + \frac{1}{2}\,\tilde{g}^{mn}\,\bm{P}_m\wedge *_\gamma \bm{P}_n + \frac{1}{2}\,\beta^{mn}\,\bm{P}_m\wedge \bm{P}_n - F_2\Bigr)\,.
\label{eq:1st-action}
\end{align}
If we identify the fields $\tilde{g}^{mn}(x)$ and $\beta^{mn}(x)$ with the generalized metric $\cH_{IJ}$ as
\begin{align}
 \cH_{IJ} = \begin{pmatrix} \tilde{g}_{mn} & -\tilde{g}_{mp}\,\beta^{pn} \\ \beta^{mp}\,\tilde{g}_{pn} & (\tilde{g}^{-1}-\beta\,\tilde{g}\,\beta)^{mn} \end{pmatrix},
\end{align}
this action reproduces the standard string action after eliminating the auxiliary fields $\bm{P}_m$\,.\footnote{The action $S_{1\text{st}}$ is related to the action $S$ \eqref{eq:DSM} (with $\Exp{\lambda}=\mu_1$) as
\begin{align*}
 S_{1\text{st}} = S + \frac{\mu_1}{4}\int_\Sigma \tilde{\cH}_{IJ}\,\bigl(DX^I -\cH^I{}_K\,*_\gamma DX^K\bigr)\wedge *_\gamma \bigl(DX^J -\cH^J{}_L\,*_\gamma DX^L\bigr)\,,
\end{align*}
where we have identified $\bm{P}_m$ with $D\tilde{x}_m$ and have defined $(\tilde{\cH}_{IJ}) \equiv \bigl(\begin{smallmatrix} 0 & 0 \\ 0 & \tilde{g}^{mn}\end{smallmatrix}\bigr)$. 
As long as $\tilde{g}^{mn}$ is invertible, the equations of motion obtained from $S$ and $S_{1\text{st}}$ are equivalent.}
The action $S_{1\text{st}}$ is not manifestly $T$-duality covariant, but as it is discussed in \cite{1502.08005}, we can manifest the symmetry as follows. 
We expand the 1-form $\bm{P}_m$ as
\begin{align}
 \bm{P}_m \equiv p_m\,\frac{N *\rmd\tau}{\sqrt{h}} + q_m\,\rmd\tau = p_m\,(\rmd\sigma +N^\sigma\,\rmd\tau) + q_m\,\rmd\tau \,,
\end{align}
and then the first-order action becomes
\begin{align}
\begin{split}
 S_{1\text{st}} &= \mu_1\int_\Sigma\rmd^2\sigma\,\Bigl(\dot{x}^m\, p_m - N^\sigma\,x'^m\,p_m - x'^m\,q_m - \frac{N}{2\sqrt{h}}\,\tilde{g}^{mn}\,p_m\,p_n
\\
 &\qquad\qquad\qquad + \frac{\sqrt{h}}{2N}\,\tilde{g}^{mn}\,q_m\,q_n + \beta^{mn}\,q_m\,p_n \Bigr)- \mu_1\int_\Sigma F_2 \,.
\end{split}
\label{eq:1st-non-covariant}
\end{align}
Eliminating the auxiliary field $q_m$\,, we obtain the Hamiltonian action
\begin{align}
 S_{\text{H}} 
 = \mu_1\int_\Sigma\rmd^2\sigma\,\bigl[ p_m\,\dot{x}^m - (\tilde{N}\,\cH_\perp + N^\sigma\,\cH_\sigma) \bigr] - \mu_1\int_\Sigma F_2 \,,
\end{align}
where we have defined $\tilde{N} \equiv N/\sqrt{\sfh}$ and
\begin{align}
 \cH_\perp \equiv \frac{1}{2}\,\cH^{IJ}\, Z_I\,Z_J \,,\quad
 \cH_\sigma \equiv \frac{1}{2}\,\eta^{IJ}\, Z_I\,Z_J \,,\quad 
 Z_I \equiv D_\sigma X_I \equiv \begin{pmatrix} p_m \\ x'^m \end{pmatrix}.
\end{align}
We note that $p_m(\sigma)$ is the usual momentum that is canonical conjugate to $x^m(\sigma)$\,.
This action can be also expressed in a $T$-duality-manifest form as
\begin{align}
 S_{\text{H}} = \mu_1 \int_\Sigma\rmd^2\sigma\,\bigl[ Z_I\,(\Pi_+^{(F)})^I{}_J\,D_\tau X^J - (\tilde{N}\,\cH_\perp + N^\sigma\,\cH_\sigma) \bigr] \,,
\end{align}
where $D_\tau X^I\equiv (\dot{x}^m,\,q_m)$ and $q_m= \tilde{g}_{mn}\,x'^n-(\tilde{g}\,\beta)_m{}^n\,p_n$\,. 
This reproduces Tseytlin action \eqref{eq:Tseytlin} under the identification $D_aX^I=\partial_aX^I$ and the conformal gauge ($\tilde{N}=1$ and $N^\sigma=0$). 
In addition, this reproduces the standard string action after eliminating $p_m$\,. 
Similar to the covariant action \eqref{eq:DSM}, the combination $(\Pi_+^{(F)})^I{}_J=\frac{1}{2}\,\eta^{IK}\,(\eta + \omega^{(F)})_{KJ}$ again plays an important role.
According to the manifest $T$-duality covariance, this Hamiltonian action can be applied to backgrounds where $\cH_{mn}$ or $\cH^{mn}$ is not invertible (see \cite{2002.12413} where it is applied to non-relativistic theories). 

\subsection{Nambu sigma model for a $p$-brane}

In the case of a general $p$-brane, a covariant action similar to Eq.~\eqref{eq:1st-action} has not been known. 
However, an extension of the action \eqref{eq:1st-non-covariant} is known as the (non-topological) Nambu sigma model \cite{1205.2595}
\begin{align}
 S&=\int_\Sigma \bigl(\rmd x^i\wedge \eta_i - \tilde{\eta}_{\bar{i}_p}\wedge\rmd x^{\bar{i}_p} - \Omega^{i\bar{j}_p}\,\tilde{\eta}_{\bar{j}_p}\wedge \eta_i
 +\tfrac{1}{2}\,\tilde{g}^{ij}\,\eta_i\wedge *_\gamma \eta_j 
 +\tfrac{1}{2}\,\tilde{g}^{\bar{i}_p\bar{j}_p}\,\tilde{\eta}_{\bar{i}_p}\wedge *_\gamma \tilde{\eta}_{\bar{j}_p}\bigr) 
\\
 &=\int_\Sigma \rmd^{p+1}\sigma\,\bigl(p_i\,\dot{x}^i - N^a\,\partial_a x^i\,p_i 
\nn\\
 &\qquad\qquad\qquad - q_{\bar{i}_p}\,\rmd_s x^{\bar{i}_p} - \Omega^{i\bar{j}_p}\, p_i\, q_{\bar{j}_p} 
 -\tfrac{N}{2\sqrt{\sfh}}\,\tilde{g}^{ij}\,p_i\,p_j 
 +\tfrac{\sqrt{\sfh}}{2N}\,\tilde{g}^{\bar{i}_p\bar{j}_p}\,q_{\bar{i}_p}\,q_{\bar{j}_p} \bigr) \,,
\label{eq:NSM}
\end{align}
where we have defined
\begin{align}
 \eta_i\equiv p_i \,\frac{N*_\gamma\rmd \tau}{\sqrt{\sfh}}\,,\qquad \tilde{\eta}_{\bar{i}_p}\equiv q_{\bar{i}_p}\,\rmd\tau\,,\qquad 
 \rmd_s x^{\bar{i}_p} \equiv \frac{\epsilon^{\tau a_1\cdots a_p}\,\partial_{a_1} x^{i_1}\cdots \partial_{a_p} x^{i_p}}{\sqrt{p!}}\,.
\end{align}
Eliminating the auxiliary fields $q_{\bar{i}_p}$\,, we obtain the Hamiltonian action
\begin{align}
 S_{\text{H}}=\int_\Sigma \rmd^{p+1}\sigma\,\bigl(p_i\,\dot{x}^i - \tilde{N}\,\cH_\perp - N^a\,\cH_a \bigr) \,,
\end{align}
where we have defined $\tilde{N} \equiv \tfrac{N}{\sqrt{\sfh}}$ and
\begin{align}
 \cH_\perp &\equiv \tfrac{1}{2}\,\cH^{IJ}\,Z_I\,Z_J\,,\qquad 
 \cH_a \equiv \partial_a x^i\,p_i \,,
\\
 \cH^{IJ} &\equiv \begin{pmatrix} \delta^i_k & \Omega^{i\bar{k}_p} \\ 0 & \delta_{\bar{i}_p}^{\bar{k}_p} \end{pmatrix} \begin{pmatrix} \tilde{g}^{kl} & 0 \\ 0 & \tilde{g}_{\bar{k}_p\bar{l}_p} \end{pmatrix} \begin{pmatrix} \delta_l^j & 0 \\ \Omega^{j\bar{l}_p} & \delta^{\bar{l}_p}_{\bar{j}_p} \end{pmatrix},\qquad 
 Z_I \equiv \begin{pmatrix} p_i \\ \rmd_s x^{\bar{i}_p} \end{pmatrix} .
\end{align}

We can consider the membrane theory by choosing $p=2$\,. 
In particular for $n\leq 4$\,, we can understand the index $I$ as that of the $R_1$-representation, and then we can express the Hamiltonian action in an $E_n$ $U$-duality-invariant form
\begin{align}
 S_{\text{H}}=\int_\Sigma \rmd^{p+1}\sigma\,\bigl[Z_I\,(\Pi_+^{(F)})^I{}_J\,D_\tau X^J - (\tilde{N}\,\cH_\perp + N^a\,\cH_a) \bigr] \,.
\end{align}
Here, in order to manifest the covariance, we have introduced a total-derivative term that contains the gauge field $F_3$ (see \cite{2002.12413} where this Hamiltonian action has been applied to non-relativistic theories). 
The point we would like to stress is that the projector $\Pi_+^{(F)}$ or the product structure $K^{(F)}$ again plays an important role, and it is a natural extension of the product structure studied in the context of the para-Hermitian geometry or the Born geometry. 
It is also noted that, for $n\geq 5$\,, this Hamiltonian action is not $U$-duality covariant, but still describes the standard membrane theory under the identification \cite{1205.2595}
\begin{align}
 \cH^{IJ} = \begin{pmatrix} \delta^i_k & 0 \\ -A_{\bar{i}_2k} & \delta_{\bar{i}_2}^{\bar{k}_2} \end{pmatrix} \begin{pmatrix} g^{kl} & 0 \\ 0 & g_{\bar{k}_2\bar{l}_2} \end{pmatrix} \begin{pmatrix} \delta_l^j & -A_{l\bar{j}_2} \\ 0 & \delta^{\bar{l}_2}_{\bar{j}_2} \end{pmatrix}.
\end{align}

Other brane theories (i.e., $p\neq 2$) can be also studied in a similar way, but the action \eqref{eq:NSM} reproduces only a part of the bosonic action. 
For example, if we consider $p=5$\,, we will obtain the M5-brane action with $A_3 = F_3 = F_6 = 0$\,. 
Thus, the action \eqref{eq:NSM} needs to be modified in order to consider the full bosonic theory. 

\section{Conclusions}
\label{sec:conclusions}

By following the recent proposals that the doubled space is naturally defined as the para-Hermitian manifold or the Born manifold, we have introduced two types of almost product structures in the exceptional space: one defines the M-theory section while the other defines the type IIB section. 
By using the almost product structures, we have defined $\omega$ in each section and proposed natural extensions of the Born sigma model. 
The obtained actions are the same as the ones studied in \cite{1607.04265,1712.10316} and reproduce the standard worldvolume theories for M2- and M5-branes as well as the $(p,q)$-string in type IIB theory. 
We have also studied the Hamiltonian actions for the string and the membrane and observed that the product structure $K^{(F)}$ again appears in the action. 

In the doubled space, the para-complex structure $K$ has played an important role in defining the physical subspace, and the section condition can be understood as the para-holomorphicity of the physical fields. 
Various mathematical structures of the doubled space have been studied in the literature, but the geometry of the exceptional space has been poorly understood. 
The analysis presented in this paper suggests that the proposed (almost) product structure $K^I{}_J$ is a natural extension of the para-complex structure in the doubled space, and it will be useful to describe the exceptional geometry in a more general framework. 
Indeed, as discussed in section \ref{sec:Lie-EFT}, under a general choice of the almost product structure $K$, we need to modify the generalized Lie derivative by using the generalized torsion associated with $K$. 
This will lead to the modifications of the generalized curvature in the exceptional space, and it might be an important future task to establish the geometry of the exceptional space by using the almost product structure. 

It is also interesting to investigate whether we can formulate the manifestly $U$-duality invariant action that reproduce all of the brane actions. 
In the present formulation, we fix the dimension of the worldvolume to be $p+1$ in advance, and it is impossible to realize other brane action with a different dimensionality by performing a $U$-duality transformation. 
However, according to the discussion given in \cite{1904.06714} (as well as the discussion of the boundary condition given in this paper), a string can end on a $N_n$-dimensional generalized Dirichlet brane\footnote{As was noted in \cite{1904.06714}, this brane may additional extend along the uncompactified external directions.} in the $D_n$-dimensional $E_n$ exceptional space, where the pair $\{N_n,\,D_n\}$ can be summarized as
\begin{align}
\begin{array}{c||ccccccc}
 n & 2 & 3 & 4 & 5 & 6 & 7 & \cdots \\\hline\hline
 N_n & 2 & 4 & 6 & 8 & 11 & 24 & \cdots \\\hline
 D_n & 3 & 6 & 10 & 16 & 27 & 56 & \cdots
\end{array}.
\end{align}
According to this proposal, when this brane has a $(p+1)$-dimensional overlap with the physical space, it is understood as the familiar $p$-brane (see \cite{hep-th:0406102} where this viewpoint was proposed in the context of the doubled space). 
Then, it might be possible to formulate the effective theory of such generalized Dirichlet brane which reproduces the standard brane actions through a certain procedure that reduces the worldvolume dimension. 
In the case of the doubled space, an effective Lagrangian that describes all of the D$p$-brane in a unified manner has been formulated in \cite{1206.6964} (see also \cite{1712.01739,1903.05601} for relevant recent works) and it is interesting to extend that to the case of the exceptional space. 

\subsection*{Acknowledgments}

The work by Y.S.\ is supported by JSPS Grant-in-Aids for Scientific Research (C) 18K13540 and (B) 18H01214. 

\appendix

\section{Notations}
\label{app:conventions}

We denote the worldvolume coordinates as $\sigma^a$ ($a=0,\dotsc,p$) and the antisymmetric symbols are defined as $\epsilon^{0\cdots p}=-\epsilon_{0\cdots p}=1$ and $\varepsilon_{a_0\cdots a_p}\equiv \sqrt{-\gamma}\,\epsilon_{a_0\cdots a_p}$\,. 
The volume form is denoted as $\rmd^{p+1}\sigma\equiv \rmd \sigma^0\wedge\cdots\wedge\rmd \sigma^p$ and the Hodge star operator is defined as $*_\gamma (\rmd \sigma^{a_1}\wedge \cdots \wedge \rmd \sigma^{a_n})\equiv \frac{1}{(p+1-n)!}\,\varepsilon^{a_1\cdots a_n}{}_{b_1\cdots b_{p+1-n}}\,\rmd \sigma^{b_1}\wedge\cdots\wedge\rmd \sigma^{b_{p+1-n}}$\,.

The coordinates on the physical subspace in the doubled space or the type IIB section of the exceptional space are denoted by $x^m$ ($m=1,\dotsc,d\equiv n-1$). 
The coordinates on the physical subspace in the M-theory section of the exceptional space are denoted by $x^i$ ($i=1,\dotsc,n$). 

The antisymmetrization is normalized such that $A_{[[i_1\cdots i_k]]}=A_{[i_1\cdots i_k]}$ is satisfied, and we define $\delta^{i_1\cdots i_k}_{j_1\cdots j_k}\equiv \delta^{[i_1}_{[j_1}\cdots \delta^{i_k]}_{j_k]}$\,.
We also define $\bm{\delta}^{i_1\cdots i_k}_{j_1\cdots j_k}\equiv k!\,\delta^{i_1\cdots i_k}_{j_1\cdots j_k}$\,. 
The usage of the multiple-index notation $\bar{i}_p$ is explained in detail at the beginning of section \ref{sec:exceptional}. 

\subsection{M-theory}
\label{app:conventions-M}

When we study M-theory, we decomposed the $E_n$ generators into the $\GL(n)$ generators $K^i{}_j$ as well as the positive-/negative-level generators
\begin{align}
 \underline{\text{positive level}}\quad \{ R^{\bar{i}_3},\,\ R^{\bar{i}_6},\, \cdots \}\,,\qquad 
 \underline{\text{negative level}}\quad \{ R_{\bar{i}_3},\,\ R_{\bar{i}_6},\, \cdots \}\,.
\end{align}
By using these, we can construct the generalized metric $\cH_{IJ}$ (with the ``natural weight'' $0$) as
\begin{align}
 \cH_{IJ} \equiv \bigl(L^\rmT\,\hat{\cH}\,L\bigr)_{IJ}\,,\qquad 
 (L^I{}_J) \equiv \Exp{A_{\bar{i}_3} R^{\bar{i}_3}} \Exp{A_{\bar{i}_6} R^{\bar{i}_6}} \cdots \,.
\end{align}
Here, $\hat{\cH}_{IJ}$ is constructed by exponentiating the $\GL(n)$ generators as
\begin{align}
 (\hat{\cH}_{IJ}) &\equiv \begin{pmatrix}
 g_{ij} & 0 & 0 \\
 0 & g^{\bar{i}_2,\bar{j}_2} & 0 & \cdots \\
 0 & 0 & g^{\bar{i}_5,\bar{j}_5} \\
 & \vdots & & \ddots \\
 \end{pmatrix} , \qquad
 g^{\bar{i}_p,\bar{j}_p} \equiv g^{i_1k_1}\cdots g^{i_pk_p}\,\delta^{j_1\cdots j_p}_{k_1\cdots k_p}\,,
\end{align}
where an overall rescaling has been done. 
The twist matrix $L^I{}_J$ is made by using the positive-level generators, whose matrix representations in the $R_1$-representation are as follows:
\begin{align}
 R^{\bar{k}_3} &\equiv \begin{pmatrix}
 0 & 0 & 0 \\
 \delta_{\bar{j}\bar{i}_2}^{\bar{k}_3} & 0 & 0 & \cdots \\
 0 & -\delta^{\bar{j}_2 \bar{k}_3}_{\bar{i}_5} & 0 \\
 & \vdots & & \ddots \\
 \end{pmatrix} , 
\qquad
 R^{\bar{k}_6} \equiv \begin{pmatrix}
 0 & 0 & 0 \\
 0 & 0 & 0 & \cdots \\
 \delta_{\bar{j}\bar{i}_5}^{\bar{k}_6} & 0 & 0 \\
 & \vdots & & \ddots \end{pmatrix},\quad \cdots\,.
\end{align}
The fields $\{g_{ij},\, A_3,\,A_6\}$ are standard bosonic fields in 11D supergravity.\footnote{The $p$-form fields $A_3$ and $A_6$ have the opposite sign compared with those used in \cite{1712.10316,1909.01335}.} 
The generalized metric $\cM_{IJ}$ with the ``weight'' $0$ is given by $\cM_{IJ}=\abs{\det (g_{ij})}^{\frac{1}{n-2}}\,\cH_{IJ}$ which is an element of the $E_n$ group and has the unit determinant.

\subsection{Type IIB theory}
\label{app:conventions-B}

When we study type IIB theory, we parameterize the generalized metric $\cH_{IJ}$ as
\begin{align}
 \cH_{IJ} \equiv \bigl(L^\rmT\,\hat{\cH}\,L\bigr)_{IJ}\,,\qquad 
 (L^I{}_J) \equiv \Exp{A^\alpha_{\bar{m}_2}\,R_\alpha^{\bar{m}_2}} \Exp{A_{\bar{m}_4}\,R^{\bar{m}_4}} \Exp{A^\alpha_{\bar{m}_6}\,R_\alpha^{\bar{m}_6}} \cdots \,,
\end{align}
where\footnote{The $p$-form fields $\{C_0,\,A^\alpha_2,\,A_4,\,A^\alpha_6\}$ have the opposite sign compared with those used in \cite{1712.10316,1909.01335}.}
\begin{align}
 (\hat{\cH}_{IJ}) &\equiv \begin{pmatrix}
 \sfg_{mn} & 0 & 0 & 0 \\
 0 & \!\!\! m_{\alpha\beta}\,\sfg^{mn}\!\!\! & 0 & 0 & \cdots \\
 0 & 0 & \sfg^{\bar{m}_3,\bar{n}_3} & 0 \\
 0 & 0 & 0 & \!\!\!m_{\alpha\beta}\,\sfg^{\bar{m}_5,\bar{n}_5}\!\!\! \\
 & \vdots & & & \ddots \\
 \end{pmatrix} , \quad \sfg^{\bar{m}_q,\bar{n}_q} \equiv \sfg^{m_1p_1}\cdots \sfg^{m_qp_q}\,\delta^{n_1\cdots n_q}_{p_1\cdots p_q}\,,
\\
 (m_{\alpha\beta}) &\equiv \Exp{\Phi}\begin{pmatrix} \Exp{-2\Phi} +(C_0)^2 & -C_0 \\ -C_0 & 1 \end{pmatrix}, \quad
 R_\gamma^{\bar{r}_2}
 \equiv \begin{pmatrix}
 0 & 0 & 0 & 0 \\
 \delta_\gamma^\alpha\,\delta_{\bar{n}\bar{m}}^{\bar{r}_2} & 0 & 0 & 0 & \cdots \\
 0 & \epsilon_{\gamma\beta}\,\delta_{\bar{m}_3}^{\bar{n}\bar{r}_2} & 0 & 0 \\
 0 & 0 & -\delta_\gamma^\alpha\,\delta_{\bar{m}_5}^{\bar{n}_3\bar{r}_2} & 0 \\
 & \vdots & & & \ddots \\
 \end{pmatrix} ,
\\
 R^{\bar{r}_4} &\equiv \begin{pmatrix}
 0 & 0 & 0 & 0 \\
 0 & 0 & 0 & 0 & \cdots \\
 \delta^{\bar{r}_4}_{\bar{n}\bar{m}_3} & 0 & 0 & 0 \\
 0 & \delta_\beta^\alpha\, \delta^{\bar{n}\bar{r}_4}_{\bar{m}_5} & 0 & 0 \\
 & \vdots & & & \ddots \\
 \end{pmatrix} ,
\qquad
 R_\gamma^{\bar{r}_6} \equiv \begin{pmatrix}
 0 & 0 & 0 & 0 \\
 0 & 0 & 0 & 0 & \cdots \\
 0 & 0 & 0 & 0 \\
 \delta_\gamma^\alpha\,\delta^{\bar{r}_6}_{\bar{n}\bar{m}_5} & 0 & 0 & 0 \\
 & \vdots & & & \ddots \\
 \end{pmatrix} .
\end{align}
The generalized metric $\cM_{IJ}$ with the unit determinant is given by $\cM_{IJ}=\abs{\det (\sfg_{mn})}^{\frac{1}{n-2}}\,\cH_{IJ}$\,.
We note that $A^\alpha_2$ can be parameterized as $(A^\alpha_2)=(B_2,\,-C_2)$ and $A_4$ is a $S$-duality-invariant combination $A_4=C_4+\frac{1}{2}\,C_2\wedge B_2$\,. 
The 6-form also can be parameterized as $(A^\alpha_6)=\bigl(C_6-\frac{1}{3!}\,C_2\wedge B_2\wedge B_2, -(B_6-\frac{2}{3!}\,B_2\wedge C_2\wedge C_2)\bigr)$ by using the standard Ramond--Ramond potential in the $C$-basis and a 6-form potential $B_6$ that couples to the NS5-brane.

\section{Charge vectors}
\label{app:charge-vector}

In this appendix, we review the construction of the charge vector $q^{\mathcal{I}}_{(\text{brane})}$ for the standard branes\footnote{The charge vector for the Kaluza--Klein monopole has been discussed in \cite{1712.10316} but those for exotic branes have never been studied.} \cite{1712.10316}. 
In the $R_2$-representation, there exists a component with $(p-1)$ antisymmetrized indices that corresponds to a $p$-brane. 
The pure charge vector $\bar{q}^{\mathcal{I}}_{(\text{brane})}$ is defined by putting $\frac{\mu_p}{p}\,\rmd x^{\bar{i}_{p-1}}$ to that component. 
For example, in M-theory the pure charge vectors for M2-brane and M5-brane are
\begin{align}
 \bar{q}^\cI_{(\text{M2})} \equiv \frac{\mu_2}{2}\begin{pmatrix} \rmd x^i \\ 0 \\ \vdots \end{pmatrix},\qquad
 \bar{q}^\cI_{(\text{M5})} \equiv \frac{\mu_5}{5}\begin{pmatrix} 0 \\ \rmd x^{\bar{i}_4} \\ \vdots \end{pmatrix}.
\end{align}
In type IIB theory, those for a $(p,q)$ string, D3-brane, and a $(p,q)$ 5-brane are
\begin{align}
 \bar{q}^\cI_{(p,q)\text{-1}} \equiv \mu_1 \begin{pmatrix} q_\alpha \\ 0 \\ 0 \\ \vdots \end{pmatrix},\qquad
 \bar{q}^\cI_{(\text{D3})} \equiv \frac{\mu_3}{3} \begin{pmatrix} 0 \\ \rmd x^{\bar{m}_2} \\ 0 \\ \vdots \end{pmatrix} , \qquad
 \bar{q}^\cI_{(p,q)\text{-5}} \equiv \frac{\mu_5}{5} \begin{pmatrix} 0 \\ 0 \\ q_\alpha\,\rmd x^{\bar{m}_4} \\ \vdots \end{pmatrix} .
\end{align}
Here, the string and the 5-brane behave as $S$-duality doublets, and we have introduced a vector $q_\alpha$, where $(q_\alpha)=(1,0)$ corresponds to the fundamental string/D5-brane while $(q_\alpha)=(0,-1)$ corresponds to the D1-brane/NS5-brane. 
These pure charge vectors do not transform covariantly under the generalized Lie derivative (i.e., under the $p$-form gauge transformations). 
In order to obtain covariant vectors, we need to multiply a twist matrix $\cL$ as follows. 

To construct the twist matrix, we need the matrix representations of the $E_n$ generators in the $R_2$-representation $(t_\alpha)_\cI{}^\cJ$\,. 
They can obtained by using the invariance of $\eta_{IJ;\cK}$
\begin{align}
 (t_\alpha)^L{}_I\, \eta_{LJ;\cK} + (t_\alpha)^L{}_J\, \eta_{IL;\cK} + \eta_{IJ;\cL}\,(t_\alpha)^\cL{}_\cK = 0 \,.
\end{align}
In the M-theory parameterization, the positive-level generators become
\begin{align}
 (R^{\bar{k}_3})^{\cI}{}_{\cJ} \equiv 
 \begin{pmatrix}
 0 & \delta^{i\bar{k}_3}_{\bar{j}_4} & \cdots \\
 0 & 0 \\
 \vdots & & \ddots 
\end{pmatrix} , \qquad
 (R^{\bar{k}_6})^{\cI}{}_{\cJ} \equiv 
 \begin{pmatrix}
 0 & 0 & \cdots \\
 0 & 0 \\
 \vdots & & \ddots 
\end{pmatrix},\quad \cdots \,.
\end{align}
In the type IIB parameterization, the positive-level generators are obtained as
\begin{align}
 &(R^{\bar{p}_2}_\gamma)^{\cI}{}_{\cJ} \equiv 
 \begin{pmatrix}
 0 & \epsilon_{\beta\gamma}\,\delta^{\bar{p}_2}_{\bar{n}_2} & 0 \\
 0 & 0 & \delta^{\bar{p}_2\bar{m}_2}_{\bar{n}_4} & \cdots \\
 0 & 0 & 0 \\
 & \vdots & & \ddots 
\end{pmatrix} , \qquad
 (R^{\bar{p}_4})^{\cI}{}_{\cJ} \equiv 
 \begin{pmatrix}
 0 & 0 & - \delta^\beta_\alpha\,\delta^{\bar{p}_4}_{\bar{n}_4} \\
 0 & 0 & 0 & \cdots \\
 0 & 0 & 0 
\\
 & \vdots & & \ddots 
\end{pmatrix},
\\
 &(R^{\bar{p}_6}_\gamma)^{\cI}{}_{\cJ} \equiv 
 \begin{pmatrix}
 0 & 0 & 0 \\
 0 & 0 & 0 & \cdots \\
 0 & 0 & 0 \\
 & \vdots & & \ddots 
\end{pmatrix} , \quad \cdots \,.
\end{align}

Now, by using these matrices, we construct the twist matrix in each theory as
\begin{align}
 \text{M-theory:}\qquad&\bigl(\cL^{\cI}{}_{\cJ}\bigr) \equiv \Exp{-F_{\bar{i}_3}\,R^{\bar{i}_3}}\Exp{-F_{\bar{i}_6}\,R^{\bar{i}_6}} \cdots\,,
\\
 \text{Type IIB theory:}\qquad&\bigl(\cL^{\cI}{}_{\cJ}\bigr) \equiv \Exp{-F^\alpha_{\bar{m}_2}\,R_\alpha^{\bar{m}_2}}\Exp{-F_{\bar{m}_4}\,R^{\bar{m}_4}}\Exp{-F^\alpha_{\bar{m}_6}\,R_\alpha^{\bar{m}_6}} \cdots\,.
\end{align}
By using the twist matrix, we construct the charge vector as
\begin{align}
 q^\cI_{(\text{brane})} \equiv \cL^{\cI}{}_{\cJ}\,q^\cJ_{(\text{brane})}\,.
\end{align}
In M-theory, we obtain
\begin{align}
 q^\cI_{(\text{M2})} \equiv \frac{\mu_2}{2}\begin{pmatrix} \rmd x^i \\ 0 \\ \vdots \end{pmatrix},\qquad
 q^\cI_{(\text{M5})} \equiv \frac{\mu_5}{5}\begin{pmatrix} -\rmd x^i\wedge F_3 \\ \rmd x^{\bar{i}_4} \\ \vdots \end{pmatrix},
\end{align}
and in type IIB theory, we obtain
\begin{align}
 q^\cI_{(p,q)\text{-1}} \equiv \mu_1{\footnotesize\begin{pmatrix} q_\alpha \\ 0 \\ 0 \\ \vdots \end{pmatrix}},\quad
 q^\cI_{(\text{D3})} \equiv \frac{\mu_3}{3}{\footnotesize\begin{pmatrix} -\epsilon_{\alpha\gamma}\,F_2^\gamma \\ \rmd x^{\bar{m}_2} \\ 0 \\ \vdots \end{pmatrix}},
\quad
 q^\cI_{(p,q)\text{-5}} \equiv \frac{\mu_5}{5} {\footnotesize\begin{pmatrix} q_\alpha\,F_4 + \frac{1}{2}\,q_\gamma\,\epsilon_{\alpha\delta}\,F^\gamma_2\wedge F_2^\delta \\ -q_\gamma\,\rmd x^{\bar{m}_2}\wedge F_2^\gamma \\ q_\alpha\,\rmd x^{\bar{m}_4} \\ \vdots \end{pmatrix}}.
\end{align}
These charge vectors transform covariantly as discussed in \cite{1712.10316}.

\end{document}